\documentclass[aps,twocolumn,superscriptaddress]{revtex4-2}

\usepackage{graphicx} 
\setkeys{Gin}{width=0.8\linewidth,totalheight=\textheight,keepaspectratio}
\graphicspath{{"figures/"}}%
\usepackage{caption}
\usepackage{enumitem,amssymb,amsmath}
\usepackage{pifont}
\usepackage{float}

\newcommand{\kir}{\text{kipa}}
\newcommand{\dc}{\text{DC}}

\newcommand{\dpa}{\text{DPA}}

\begin{document}

\title{A near-ideal degenerate parametric amplifier}
\author{Daniel J. Parker}
\affiliation{School of Electrical Engineering and Telecommunications, UNSW Sydney, Sydney, NSW 2052, Australia}
\author{Mykhailo Savytskyi}
\affiliation{School of Electrical Engineering and Telecommunications, UNSW Sydney, Sydney, NSW 2052, Australia}
\author{Wyatt Vine}
\affiliation{School of Electrical Engineering and Telecommunications, UNSW Sydney, Sydney, NSW 2052, Australia}
\author{Arne Laucht}
\affiliation{School of Electrical Engineering and Telecommunications, UNSW Sydney, Sydney, NSW 2052, Australia}
\author{Timothy Duty}
\affiliation{School of Physics, UNSW Sydney, Sydney, NSW 2052, Australia}
\author{Andrea Morello}
\affiliation{School of Electrical Engineering and Telecommunications, UNSW Sydney, Sydney, NSW 2052, Australia}
\author{Arne L. Grimsmo}
\affiliation{Centre for Engineered Quantum Systems, School of Physics, The University of Sydney, Sydney, Australia}
\author{Jarryd J. Pla}
\affiliation{School of Electrical Engineering and Telecommunications, UNSW Sydney, Sydney, NSW 2052, Australia}


\begin{abstract}
Degenerate parametric amplifiers (DPAs) exhibit the unique property of phase-sensitive gain and can be used to noiselessly amplify small signals or squeeze field fluctuations beneath the vacuum level. In the microwave domain, these amplifiers have been utilized to measure qubits in elementary quantum processors, search for dark matter, facilitate high-sensitivity spin resonance spectroscopy and have even been proposed as the building blocks for a measurement based quantum computer. Until now, microwave DPAs have almost exclusively been made from nonlinear Josephson junctions, which exhibit high-order nonlinearities that limit their dynamic range and squeezing potential. In this work we investigate a new microwave DPA that exploits a nonlinearity engineered from kinetic inductance. The device has a simple design and displays a dynamic range that is four orders of magnitude greater than state-of-the-art Josephson DPAs. We measure phase sensitive gains up to $50\,\text{dB}$ and demonstrate a near-quantum-limited noise performance. Additionally, we show that the higher-order nonlinearities that limit other microwave DPAs are almost non-existent for this amplifier, which allows us to demonstrate its exceptional squeezing potential by measuring the deamplification of coherent states by as much as 26~dB.
\end{abstract}

\maketitle

\section{Introduction}
High performance cryogenic microwave amplifiers have become crucial components for an increasing number of contemporary experiments in condensed matter physics and quantum engineering. Microwave amplifiers that are based on parametric photon conversion are particularly promising since they can operate at the quantum-noise-limit, where only the minimal amount of noise required by quantum mechanics is added to the amplified signal. These amplifiers have facilitated the high-fidelity readout of quantum bits in elementary quantum processors \cite{vijay_2011}, enabled spin resonance spectroscopy of femtolitre-volume samples \cite{ranjan_2020} and are even aiding the search for axions \cite{malnou_2019,backes_2021}.

Parametric amplifiers largely fall into one of two classes: phase insensitive or phase sensitive. In quantum mechanics, an electromagnetic field $X = X_1\cos(\omega t) + X_2\sin(\omega t)$ (with angular frequency $\omega$) can be described by dimensionless quadrature field operators $\hat{X_1} = (\hat{a}^\dagger+\hat{a})/2$ and $\hat{X_2} = i(\hat{a}^\dagger-\hat{a})/2$, where $\hat{a}^\dagger$ and $\hat{a}$ are the boson annihilation and creation operators. A phase insensitive amplifier applies gain $G$ equally to both quadratures $\langle\hat{Y_1}\rangle=G\langle\hat{X_1}\rangle$ and $\langle\hat{Y_2}\rangle=G\langle\hat{X_2}\rangle$ (where $\hat{Y_1}$ and $\hat{Y_2}$ represent the field at the output of the amplifier), unavoidably adding at least $1/4$ photon of noise to each quadrature in the process. Conversely for a phase-sensitive amplifier, one field quadrature is amplified $\langle\hat{Y_1}\rangle=G\langle\hat{X_1}\rangle$, whilst the other is deamplified $\langle\hat{Y_2}\rangle=\langle\hat{X_2}\rangle/G$. This allows for amplification of a single quadrature without any added noise \cite{caves_1982}. The noiseless nature of a phase-sensitive amplifier makes it distinctly useful for detecting small microwave signals, particularly those at the single photon level \cite{kindel_2016}.

In addition to its superior noise performance, a phase-sensitive amplifier can be used to reduce the fluctuations of an electromagnetic field. A mode of electromagnetic radiation cooled to its ground state will exhibit a quantum mechanical noise referred to as `vacuum fluctuations'. These fluctuations obey the uncertainty relation $\delta\hat{X_1^2}\delta\hat{X_2^2} \ge 1/16$, where $\delta\hat{X_1^2}$ and $\delta\hat{X_2^2}$ represent the variances of the field quadratures (in dimensionless units of photons), and establish the ultimate limit to noise for measurements of an electromagnetic field. When a field in its quantum ground state enters a phase-sensitive amplifier, the vacuum fluctuations are deamplified or `squeezed' along one quadrature at the expense of increasing them along the other. Squeezed noise can be used to enhance the signal-to-noise ratio (SNR) in measurements and has been successfully deployed, for example, in gravitational wave detection \cite{aasi_2013}.

Squeezed vacuum states are also a valuable resource in quantum computing \cite{menicucci_2006}. Measurement-based computation using highly-entangled cluster states encoded in the modes of an electromagnetic field is one credible pathway to achieving large-scale quantum computation \cite{asavanant_2019}. Critically, it has been shown that fault-tolerance in this scheme can be attained using vacuum states squeezed by at least $15-17\,\text{dB}$~\cite{walshe_2019}. Circuit-based microwave squeezers are a particularly attractive platform in this context, as they combine circuit manufacturability with another key requirement in cluster-state computing; the ability to engineer non-Gaussian states of light \cite{hofheinz_2008,grimsmo_2017,grimm_2020}.

\begin{figure*}
  \includegraphics[width=172mm]{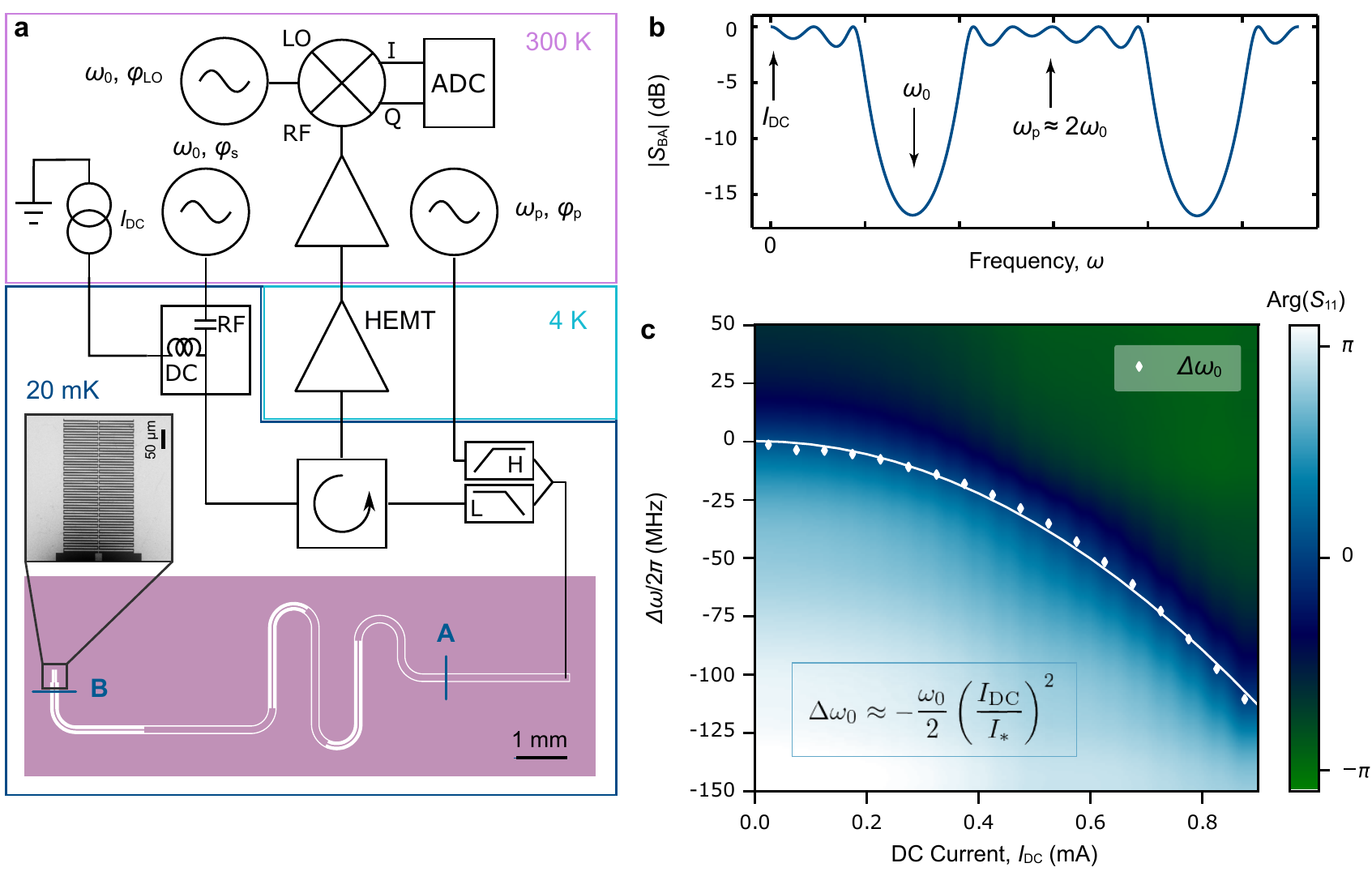}
  \caption{\label{fig1}
     {\bf Experimental setup and device geometry.}
     {\bf a} Experimental setup, showing the device pattern (bottom) and close-up
             optical microscope image of the interdigitated capacitor (IDC). The signal tone and bias current are combined at $20\,\text{mK}$ using a bias tee. A diplexer mixes the pump and signal tones immediately before the sample. A circulator splits the forward microwave path from the reflected output of the device, which is measured at room temperature using homodyne detection (depicted) or with a vector network analyzer.
     {\bf b} Simulated frequency response of the two port bandstop filter, with
             ports $A$ and $B$ as marked on the pattern in {\bf a}. The filter
             passes DC and $2\omega_0$, and attenuates $\omega_0$. The filter connects at port $B$ to a quarter-wavelength resonator with resonant frequency $\omega_0$.
     {\bf c} The phase response of the device vs bias current. The bias current
             shifts the resonant frequency over more than $100$~MHz. Measured
             resonant frequencies at each current are presented (white
             diamonds), alongside a fit to the inset equation (solid white line).
  }
\end{figure*}

In the microwave domain, the Josephson Parametric Amplifier (JPA) represents the state-of-the-art in phase-sensitive amplification technology. A JPA consists of one or more Josephson junction, typically in the form of a SQUID loop, embedded in a low quality factor superconducting resonator \cite{yurke_1989,roy_2016}. Vacuum squeezing has been achieved with JPAs employing single cavity modes (so-called degenerate parametric amplifiers) at the level of $10\, \text{dB}$ \cite{castellanos-beltran_2008} and through the entanglement of two distinct cavity modes ($12\,\text{dB}$) \cite{eichler_entanglement_2014}. However, recent experimental \cite{bienfait_2017,malnou_2018} and theoretical \cite{boutin_2017} investigations of JPAs have uncovered differences between the JPA and an ideal degenerate parametric amplifier, which become significant in the high gain limit ($> 10\,\text{dB}$) and constrain the dynamic range and amount of achievable squeezing. Higher-order nonlinearities originating from the physics of Josephson junctions limit the useful linear regime of operation, with typical $1 \,\text{dB}$-compression points measuring less than $-90 \,\text{dBm}$ at the amplifier output \cite{zhou_2014, planat_2019,grebel_2021,grebel_2021}. Attention has recently been focused on engineering JPAs with Hamiltonians that more closely resemble that of an ideal DPA, such as those employing junctions arranged in a `SNAIL' configuration \cite{frattini_2018,sivak_2019,sivak_2020,chien_2020}, which has been successful in pushing $1 \,\text{dB}$-compression point output powers to as high as $-73 \,\text{dBm}$.

In this work we present a new type of phase-sensitive microwave parametric amplifier that behaves as a near-ideal DPA. The device contains no Josephson junctions (making it robust to electrostatic discharge) and is produced with a single-step lithography process on a thin film of NbTiN. The nonlinearity responsible for parametric amplification in this device originates from a kinetic inductance intrinsic to the NbTiN film \cite{tholen_2009,eom_2012,vissers_2016,chaudhuri_2017,anferov_2020,malnou_2021}. We observe up to $50\,\text{dB}$ of phase sensitive gain with a gain-bandwidth product of $53(7)\,\text{MHz}$. An exceptionally large $1 \,\text{dB}$-compression point output power of $-49.5(8)\,\text{dBm}$ is measured and represents a 3-5 orders of magnitude improvement over comparable JPAs.

We demonstrate that the weak nonlinearity of our amplifier has the potential to produce highly-squeezed microwave fields. Through mapping the squeezing transformation of our DPA using coherent tones, we observe deamplification levels approaching $30\,\text{dB}$ without the distortions commonly observed in JPAs \cite{bienfait_2017,boutin_2017,malnou_2018}. Finally, we explore the noise properties of our amplifier and find that it operates close to the quantum noise limit.

\section{The Kinetic Inductance Amplifier }
Kinetic inductance is associated with the energy stored in the motion of charge-carrying particles. For superconducting films, Ginzburg-Landau theory predicts a current-dependence of the kinetic inductance described by \cite{annunziata_2010, zmuidzinas_2012}:
\begin{equation}
  L_k(I) \approx L_0 \bigg[1 +
                     \bigg(\frac{I}{I_{*}}\bigg)^2
                    \bigg]
  \label{eqn1}
\end{equation}
where $L_0$ is the per-unit-length kinetic inductance of the material without a current, and $I_{*}$ determines the strength of the current dependence and is proportional to the critical current $I_c$ of the film. This form of nonlinear inductance is analogous to an optical Kerr medium. When a current passing through the film consists of two different microwave tones, i.e. a signal tone (at a frequency $\omega_s$) and a much stronger `pump' tone (at $\omega_p)$, the nonlinearity gives rise to four wave mixing (4WM), where energy transfer from the pump to the signal can produce parametric amplification \cite{eom_2012,chaudhuri_2017}. In this process, two pump photons are converted to a signal photon and a photon at an additional tone called the `idler' (with frequency $\omega_i$), where energy conservation requires $2\omega_p = \omega_s + \omega_i$. Introducing a DC current bias on top of the microwave tones $I = I_\dc + I_{\mu\text{w}}$ lowers the order of the nonlinearity:
\begin{equation}
  L_k(I) \approx L_0 \bigg[1 +
                     \bigg(\frac{I_\dc}{I_{*}}\bigg)^2 +
                     2\frac{I_\dc I_{\mu\text{w}}}{I_{*}^2} +
                     \bigg(\frac{I_{\mu\text{w}}}{I_{*}}\bigg)^2
                     \bigg]
\end{equation}

In addition to the Kerr component ($\propto I_{\mu\text{w}}^2$), a new term linear in $I_{\mu\text{w}}$ appears which can facilitate a three wave mixing (3WM) process. Here a single pump photon produces a signal photon and an idler photon, such that $\omega_p = \omega_s + \omega_i$. Three wave mixing is advantageous in the context of parametric amplification since there can be a large spectral separation between the pump and the signal. This means that the strong pump tone can be readily removed through filtering, preventing the saturation of any following amplifiers in the detection chain. 3WM-type parametric amplifiers using kinetic inductance have been demonstrated recently in traveling wave geometries \cite{vissers_2016,malnou_2021}. However, at high pump powers a competition between 4WM and 3WM processes is known to degrade the parametric gain in devices and limit their performance \cite{erickson_2017}. In addition, there has been limited experimental work on phase sensitive amplification in quantum-limited microwave travelling wave devices.

Here we implement a resonant 3WM-type DPA that utilizes kinetic inductance and exhibits phase sensitive gain. Critically, the resonant nature of our kinetic inductance parametric amplifier (KIPA) strongly suppresses 4WM and other higher-order processes, permitting extremely high levels of pure 3WM gain. The device (see Fig.~\ref{fig1}a) is fabricated in a $9.5\,\text{nm}$ thick film of NbTiN on silicon, benefiting from the high magnetic field resilience (up to $B_\perp \approx 350\,\text{mT}$) and high critical temperature ($T_c \approx 10.5\,\text{K}$) that are characteristic of this superconductor \cite{samkharadze_2016,zhang_2015}. NbTiN on silicon can exhibit extremely low losses with internal quality factors $Q_i$ greater than $10^6$ \cite{bruno_2015}, which is crucial to the generation of highly squeezed states. The amplifier is measured at a temperature of $20$~mK in a $^3$He/$^4$He dilution refrigerator (see Supplemental Materials for details).

The KIPA is defined geometrically by a coplanar waveguide (CPW) quarter-wavelength resonator coupled to a single port via a microwave Bragg mirror \cite{yun_1999,sigillito_2017}, which can equivalently be viewed as a stepped-impedance band-stop filter. The filter (which has a frequency response depicted in Fig.~\ref{fig1}b) mimics the role of a capacitive coupling element commonly found in JPAs
\cite{castellanos-beltran_2008,eichler_entanglement_2014,zhou_2014}, but importantly does not break the inner track of the CPW, allowing a DC current to pass through the device. The resonator is realised using a segment of CPW featuring an interdigitated capacitor (IDC) (see Fig.~\ref{fig1}a) terminated in a short circuit, and is designed to produce a resonance at the centre of the band-stop region $\omega_0/2\pi \approx 7.2\,\text{GHz}$. The small CPW track width in the resonator ($w = 2~\mu$m) reduces $I_*$ and provides a sizable total kinetic inductance of $L_T = 3.84\,\text{nH}$. The IDC adds additional capacitance to the resonator to decrease its characteristic impedance ($Z_0 \approx 118~\Omega$) and subsequently enhance the pump current for a given pump power, which helps to minimise device heating. Furthermore, the IDC introduces dispersion to the resonator \cite{malnou_2021}, detuning the higher-order modes away from harmonics of the fundamental (i.e. $3\omega_0$), preventing inter-mode coupling induced by the strong parametric pump \cite{zakka_2011}. The KIPA functions in the highly over-coupled regime, where the coupling rate to the port $\kappa$ far exceeds the internal rate of loss $\gamma$.

\begin{figure*}
  \includegraphics[width=172mm]{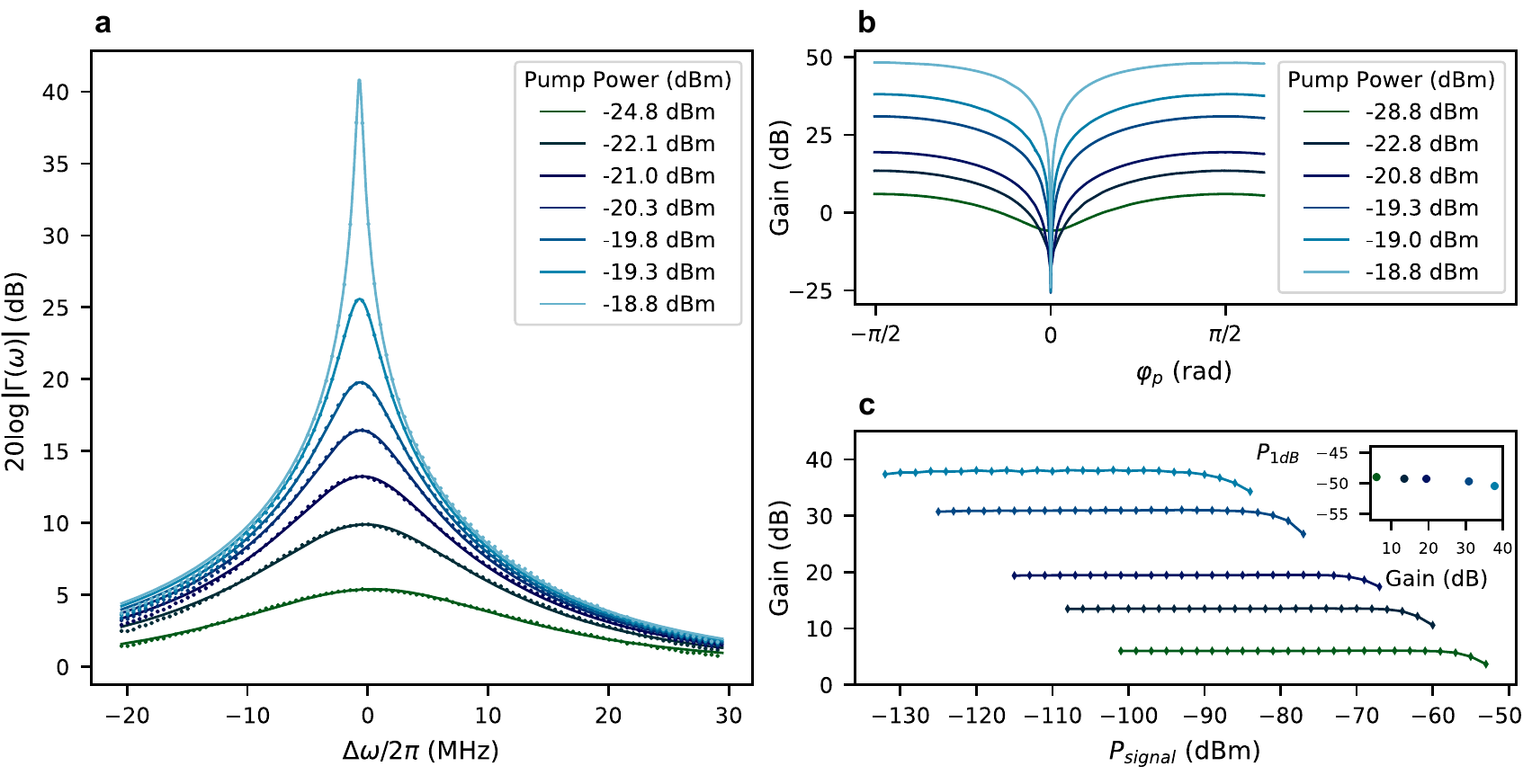}
  \caption{\label{fig2}
           {\bf Amplifier gain characteristics and compression with signal power.}
           {\bf a} Phase insensitive gain as a function of the signal frequency
                   $\omega = \Delta \omega + \omega_p/2$ for different pump powers (circles). The fitted theoretical frequency response is plotted (solid lines) with fitting parameters provided in the Supplemental Materials.
           {\bf b} Phase sensitive gain as a function of pump phase $\varphi_p$
                   for various pump powers. The phase response has been aligned such
                   that $\varphi_p = \pi/2$ corresponds to maximum gain.
           {\bf c} Peak gain (as illustrated in {\bf b}) as a function of signal
                   power for the same pump powers shown in {\bf b}. Inset: output $1\,\text{dB}$-compression power as a function of gain.
          }
\end{figure*}

To operate the KIPA, we feed the combined bias current $I_\dc$, signal and pump into its port, as illustrated in Fig.~\ref{fig1}a. The tones mix inside the resonator and the resulting amplified and reflected signal is routed to a High Electron Mobility Transistor (HEMT) amplifier at $4\,\text{K}$. This is followed by a third low-noise amplification stage at room temperature before being measured with a Vector Network Analyzer (VNA) or undergoing homodyne detection (see Supplemental Materials for details).

 We have derived the Hamiltonian for the KIPA in the presence of the bias current $I_\dc$ and a pump tone $I_p\cos(\omega_p t + \varphi_p)$, expressed in a reference frame rotating at $\omega_p/2$:
\begin{equation}
  H_{\kir}/\hbar = \underbrace{\Delta \hat{a}^\dagger\hat{a} + \frac{\xi}{2} \hat{a}{^\dagger}^2 + \frac{\xi^*}{2} \hat{a}^2}_{H_{\dpa}/\hbar } +
      \underbrace{\frac{K}{2} \hat{a}{^\dagger}^2\hat{a}^2}_{H_{\text{Kerr}}/\hbar }
  \label{eqn2}
\end{equation}
where $\Delta$ accounts for a frequency detuning of the KIPA from half the pump frequency $\omega_p/2$. $H_{\dpa}$ is the Hamiltonian for an ideal DPA \cite{boutin_2017}; it is quadratic in the field operators and is characterized by the 3WM strength $\xi$. $H_{\text{Kerr}}$ represents the next higher-order term, which here is a self-Kerr interaction with strength $K$. See the Supplemental Materials for a detailed derivation of the Hamiltonian. Eq.~\ref{eqn2} is also the same approximate Hamiltonian found for 3WM-type JPAs, such as those that employ flux-pumped SQUIDs \cite{yamamoto_2008,zhou_2014,grebel_2021} or SNAILs \cite{frattini_2018,sivak_2019,sivak_2020}. An important quantity that largely determines the dynamic range and squeezing potential in these DPAs is the ratio $\kappa/|K|$, which quantifies the relative strength of $H_{\text{Kerr}}$ to $H_{\dpa}$ \cite{boutin_2017} (since $|\xi| \rightarrow \kappa/2$ at large gain).

To quantify the Kerr interaction $|K| \approx \omega_0(\hbar \omega_0)/(I_*^2L_T)$ for the KIPA (refer to Supplemental Materials), we measure $I_*$. The DC bias current is swept in the absence of a pump tone, resulting in a shift of the device's resonance frequency (detected via its phase response, see Fig.~\ref{fig1}c) that is related to the change in kinetic inductance described by Eq.~\ref{eqn1}. We observe a resonance frequency shift of $\sim100\,\text{MHz}$ for a $0.9\,\text{mA}$ bias, and extract $I_* = 5.10(9)\,\text{mA}$ from a fit of the current dependence. We estimate the Kerr constant for this device to be $|K|/2\pi \approx 0.13\,\text{Hz}$, a completely negligible quantity relative to all other system parameters. We note that $I_*$ is about three orders of magnitude greater than the critical current of a typical JPA, indicating a much weaker form of nonlinearity. This provides a ratio $\kappa/|K| > 10^8$ for the KIPA that is several orders of magnitude greater than that of a JPA \cite{zhou_2014, planat_2019,grebel_2021}.

We expect the 3WM strength $|\xi| \approx \omega_0(I_\dc I_p/I_*^2)$ (see Supplemental Materials) to be linear in the applied DC current. For the remainder of the paper, a bias current of $I_\dc = 0.834\,\text{mA}$ is used; close to the critical current of the film but leaving a sufficient margin for additional microwave currents applied through the pump and signal. A comparison of the expressions for $|K|$ and $|\xi|$ reveals why the KIPA functions as an ideal DPA: the photon energy is a minuscule fraction of the characteristic Kerr energy (i.e. $\hbar\omega_0/(L_TI_*^2) \ll 1$) by virtue of $I_*$ being large. In fact, it can be shown that the Kerr energy is related to the superconducting pairing energy $E_p = L_TI_*^2$ \cite{zmuidzinas_2012}, which itself depends on the effective volume of the nonlinear inductance. The greater the volume over which $L_T$ is spread, the smaller the Kerr interaction. This is also observed for JPAs, where it is known that a large array of weakly-nonlinear SQUIDs distributed throughout the resonator can substantially lower the self Kerr interaction strength relative to the case of a single strongly nonlinear SQUID \cite{eichler_2014_control}. The 3WM strength for the KIPA is, on the other hand, somewhat independent of $I_*$, since $I_\dc$ and $I_p$ can always be raised to a sizable fraction of $I_*$, with the provision that they are kept sufficiently small so that heating of the refrigerator and device does not occur.

\section{Amplifier Characteristics}
\begin{figure*}
  \includegraphics[width=172mm]{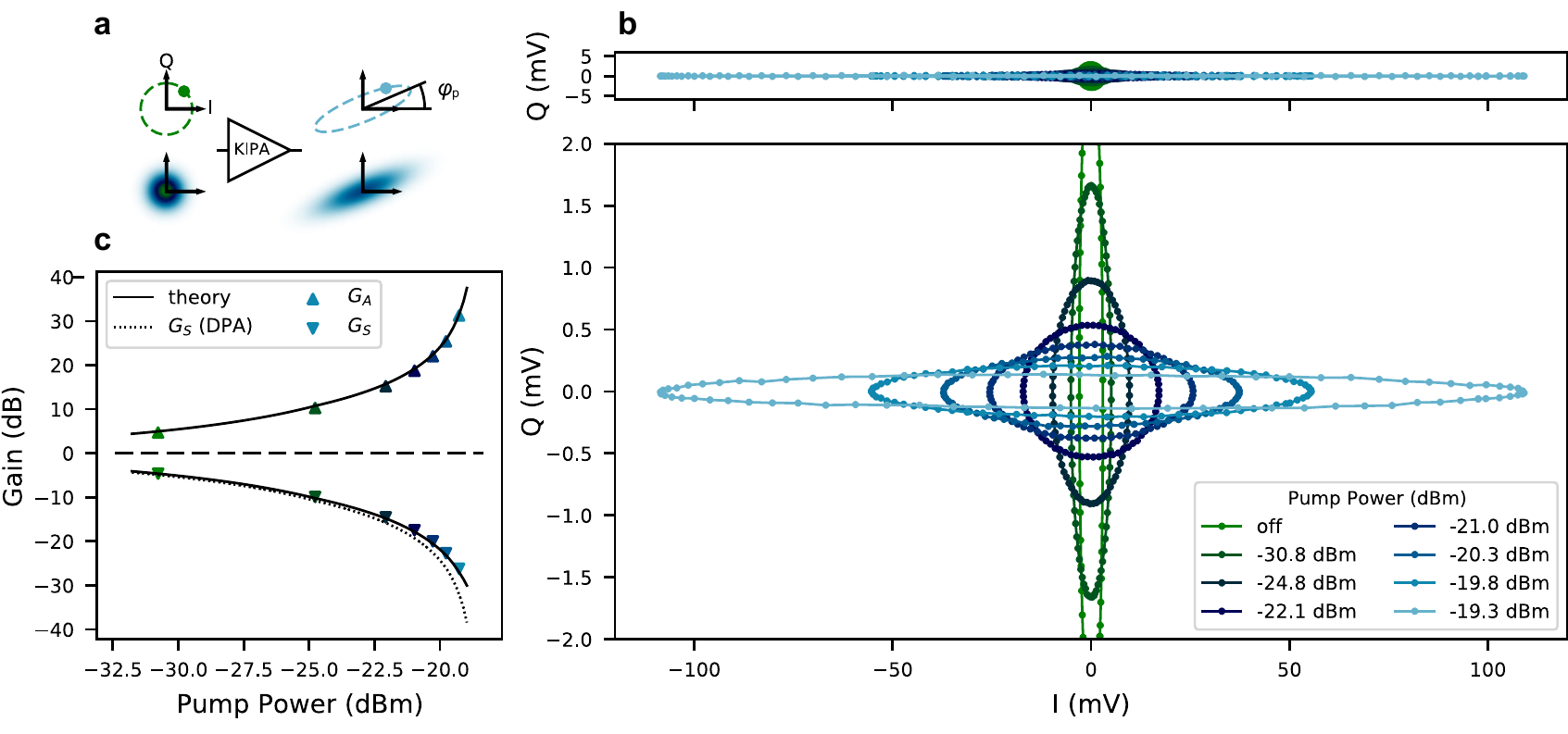}
  \caption{\label{fig3}
           {\bf The squeezing transformation, measured with coherent inputs.}
           {\bf a} An illustration of the squeezing transformation that
                   similarly distorts the IQ plane for coherent inputs (top)
                   and the quasi-probability distribution of a quantum (vacuum) state
                   (bottom).
           {\bf b} KIPA response to coherent inputs of constant amplitude and varying phase, plotted on two different scales. Top: outputs are plotted with
                   an equal aspect ratio, where the reflected input sweep with the KIPA off is observed as a circle (green) at the center of the plot. Turning the KIPA on stretches the circle to an ellipse, which resembles a blue line in this plot. Bottom: the same outputs
                   plotted with a zoomed-in scale along Q so that the
                   elliptical transformation may be observed. Solid lines are
                   a guide for the eye.
           {\bf c} The deamplification $G_S$ and amplification $G_A$ as a function
                   of pump power. Points are extracted from the ellipses {\bf b}. The dotted line is the expected deamplification for an ideal DPA and the solid lines are the amplification and deamplification calculated with a model that includes small reflections along the measurement lines (see Supplemental Materials).
          }
\end{figure*}

Applying a pump tone at the frequency $\omega_p/2\pi = 14.381\,\text{GHz}$ and bias current $I_\dc = 0.834\,\text{mA}$ produces an amplification feature centred around $\omega_0/2\pi \approx 7.1905\,\text{GHz}$ (see Fig.~\ref{fig2}a). The KIPA generates an amplified signal tone $\omega_s$ at its output, along with an idler at $\omega_i$ such that energy is conserved in the 3WM process $\omega_p = \omega_s + \omega_i$. The phase insensitive gain, which occurs when the signal and idler tones are at distinct frequencies, i.e. for $\omega_s = \omega_p/2 + \Delta\omega$ (with $|\Delta\omega|$ exceeding the resolution bandwidth of the measurement), increases with the pump power and is found to be in excess of $40\,\text{dB}$ before the KIPA crosses the threshold where spontaneous parametric oscillations occur (see Supplemental Materials). To characterise the line-shape of the amplification feature, we assume that the KIPA operates as an ideal DPA (i.e., we assume $K = 0$) and derive the reflection parameter $\Gamma$ using input-output theory \cite{boutin_2017}:

\begin{equation}
    \Gamma(\omega) = 
            \frac{\kappa(\kappa + \gamma)/2 + i\kappa(\Delta + \omega - \omega_p/2)} {\Delta^2 + \big[(\kappa + \gamma)/2 + i(\omega - \omega_p/2)\big]^2 - |\xi|^2} - 1
  \label{eqn:S11},
\end{equation}
(see Supplemental Materials for the derivation). We fit the gain data with the reflection model and observe excellent agreement with theory (see Fig.~\ref{fig2}a). We extract an average coupling quality factor $Q_c = 135$, along with a constant gain-bandwidth-product for the KIPA of $53(7)$~MHz.

When applying a signal tone at half the pump frequency $\omega_s = \omega_p/2$, the KIPA enters the degenerate mode of operation, producing phase sensitive gain as the signal and idler tones interfere. Fig.~\ref{fig2}b depicts the gain of the KIPA as a function of pump phase, where up to $26\,\text{dB}$ of deamplification and close to $50\,\text{dB}$ of amplification are measured. Compared to phase insensitive amplification, additional gain is obtained in degenerate mode due to the constructive interference that occurs between the signal and idler.

After calibrating the phase of the pump to achieve maximum amplification (i.e. $\varphi_p \approx \pi/2$), we characterise the degenerate $1\,\text{dB}$-compression point of the KIPA as a function of gain (see Fig.~\ref{fig2}c). For $\sim 20\,\text{dB}$ of phase sensitive gain, we find a 1~dB compression power of $-69.5(8)\,\text{dBm}$ at the device input, comparable to the performance of kinetic inductance travelling wave amplifiers \cite{ho_eom_2012,vissers_2016,malnou_2021}, despite the KIPA's resonant nature. The output power of the KIPA for this measurement was close to the input power $1\,\text{dB}$-compression point of the cryogenic HEMT amplifier ($\sim-46\,\text{dBm}$). It is thus possible that the true $1\,\text{dB}$-compression point of the KIPA is even higher than we report here.

\section{The Squeezing Transformation}
The phase-dependent interference of the signal and idler fields in a DPA results in an affine transformation applied to the IQ-plane of the input field, also commonly called the squeezing transformation \cite{caves_1982}. Fig.~\ref{fig3}a illustrates the distortion of the IQ-plane for coherent inputs that lie along a contour of constant amplitude, in addition to an equivalent transformation of a phase space representation of a vacuum state input. The fields, which initially occupy a circular region on the IQ-plane, are stretched to form an ellipse, with the area being conserved in the process. Coherent states are useful for studying the squeezing transformation since any noise field that couples into the cavity through the loss channel ($\gamma$) may be neglected (averaged away in a measurement), permitting a clear inspection of any deviations from the expected transformation. Hamiltonian non-idealities manifest as an S-shaped distortion of the phase space at high gains, as has been experimentally observed \cite{bienfait_2017, malnou_2018} and modelled \cite{boutin_2017} in JPAs for gains typically exceeding $\sim10\,\text{dB}$. 

\begin{figure}
  \includegraphics[width=86mm]{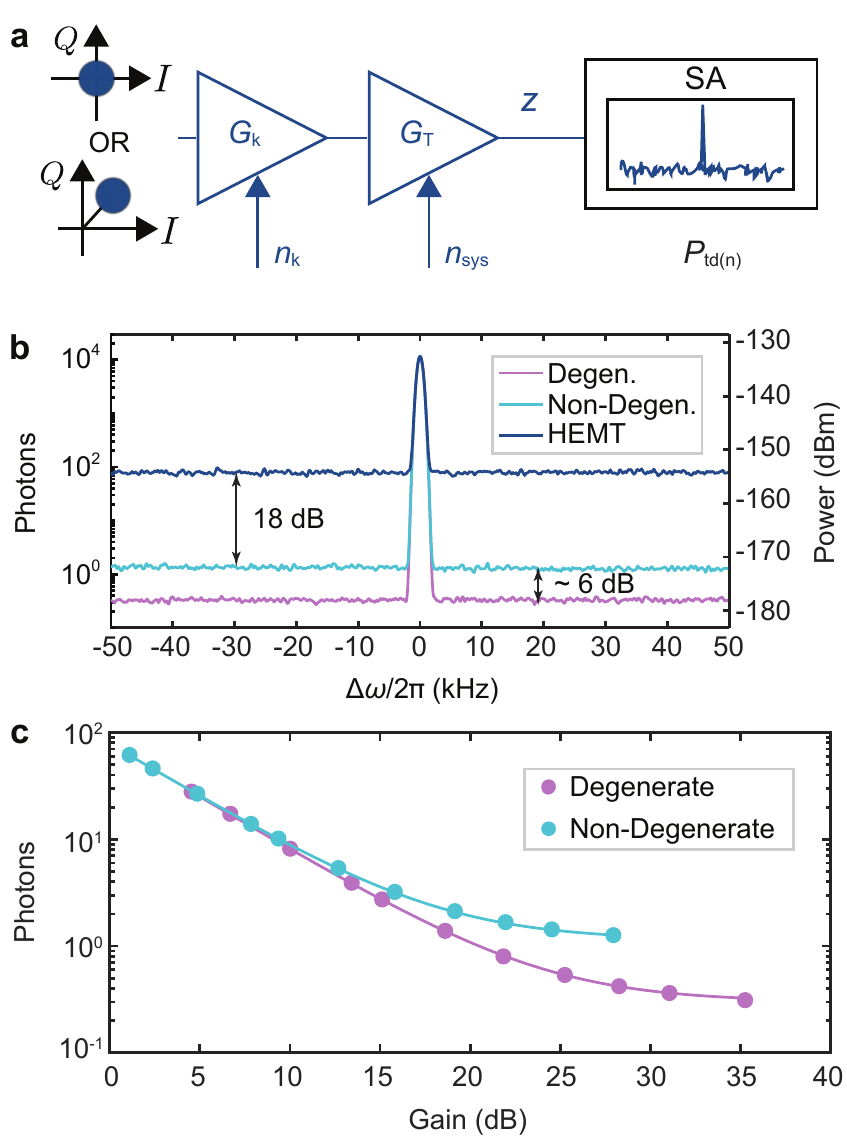}
  \caption{\label{fig4}
           {\bf Noise temperature characterisation.}
           {\bf a} Schematic depicting a simplified detection chain that can be used to model the noise properties of the KIPA. The KIPA amplifies with a gain $G_k$, adding an amount of noise $n_k$. The following amplifiers and components are modeled as a single amplifier with a total gain of $G_T$ and a noise contribution $n_\text{sys}$.
           {\bf b} Measured noise in the presence of a coherent tone, shown in units of power (dBm) and photons, referred to the input of the KIPA. The measurement is taken with a bandwidth resolution of 1~kHz. Traces were recorded with the KIPA in degenerate mode, non-degenerate mode and off. In non-degenerate mode the signal-to-noise ratio is improved by 18~dB compared to a measurement using the cryogenic HEMT as the first amplifier. An additional 6~dB enhancement is observed when the KIPA is in the degenerate mode.
           {\bf c} Number of input-referred noise photons calculated by measuring the noise power at the output of the chain over a 1~kHz bandwidth around $\omega_0$ (500~kHz detuned for the non-degenerate data), and dividing the result by $zG_TG_k$. Error bars are within the size of the markers on this scale.
          }
\end{figure}

We probe the squeezing transformation by sweeping the phase of a coherent state incident on the KIPA and use homodyne detection to measure the transformed output. A coherent tone power of $-112$~dBm (corresponding to $\sim 16$ intracavity photons) was chosen to provide sufficient signal-to-noise ratio without saturating the amplifier chain. See Supplemental Materials for details. Our results are shown in Fig.~\ref{fig3}b for different pump powers. When the KIPA is off, sweeping the phase of the input coherent state traces out a circle on the IQ-plane. Activating the KIPA maps the circle to an ellipse at the detector with no noticeable S-type distortion, even for a degenerate gain of $30\,\text{dB}$. This exceeds the achievable phase-sensitive gain without distortion observed in JPAs by approximately two orders of magnitude. Further increases in gain (up to $50\,\text{dB}$) did not produce any obvious distortions, though at these higher gains the signal power had to be reduced to avoid saturating the cryogenic HEMT and room temperature amplifiers, resulting in significant degradation in the SNR. The lack of S-type features at high gain support the conclusion that the self-Kerr correction is negligible and to an excellent approximation the KIPA can be described as an ideal DPA.

The deamplification level $G_S$ is defined as the greatest reduction in amplitude of an input coherent state by the squeezing transformation. We additionally define the amplification level $G_A$ as being the corresponding increase in gain that occurs orthogonal to the axis of deamplification. $G_S$ and $G_A$ are measured after aligning the amplification and deamplification axes along I and Q, respectively, and averaging multiple measurements (refer to Supplemental Materials for details). Fig.~\ref{fig3}c presents our results, where up to $26\,\text{dB}$ of deamplification is observed for $30\,\text{dB}$ of amplification. Using the input-output theory for an ideal DPA, we derive the squeezing transformation and predict $G_S$ and $G_A$ for the DPA parameters obtained from the fits in Fig.~\ref{fig2}a (see Supplemental Materials). We observe some asymmetry $G_S \neq G_A$ in our data at high gains, as shown in Fig.~\ref{fig3}c, which we attribute to a small $\sim$~2\% reflection occurring at the input to the KIPA (see Supplemental Materials). This reflected signal adds to the deamplified coherent state and thus reduces the measured deamplification level. Such a reflection would occur for a $2~\Omega$ impedance mismatch of the printed circuit board to which the amplifier is connected, which is within its fabrication tolerances. Including a weak reflection along the signal path in our model accurately reproduces the slight asymmetry, as shown in Fig.~\ref{fig3}c.

\section{Noise Properties}
We examine the noise performance of the KIPA by monitoring the output power of the setup on a spectrum analyzer. Fig.~\ref{fig4}a portrays a simplified noise model of the detection chain, with the KIPA amplifying either noise (a vacuum state) or a coherent state (i.e. a displaced vacuum). The following amplifiers and any loss along the output line are modeled by a single amplifier with a gain of $G_T$ (refer to Supplemental Materials). The parameters $n_k$ and $n_\text{sys}$ represent the number of noise photons in excess of the quantum limit added by each amplifier. A conversion factor $z$ translates the dimensionless units of photons to an equivalent power in watts recorded on the spectrum analyzer. The measured output power can thus be converted to a number of photons referred to the output of the KIPA by dividing by the factor $zG_T$, which is extracted from a detailed analysis of the output noise spectrum as a function of gain and temperature (see Supplemental Materials). We derive expressions for the input-referred number of noise photons of the KIPA in degenerate ($n_\text{td}$) and non-degenerate ($n_\text{tn}$) modes using input-output theory (Supplemental Materials):\\
\begin{align}
        n_\text{td} &= \frac{P_{td}}{zG_T G_k} \approx \frac{1}{4}(2n_\text{th} + 2n_{kd} + 1) + \frac{n_\text{sys}}{G_k} \label{eqn:ntd}\\
        n_{tn} &= \frac{P_{tn}}{zG_T G_k} \approx (2n_{\text{th}} + n_{kn} + 1) + \frac{n_\text{sys}}{G_k} \label{eqn:ntnd}
\end{align}\\
with $P_\text{td(tn)}$ the power at the output of the detection chain measured in degenerate (non-degenerate) mode, $n_\text{th}$ the number of thermal photons occupying the input field and $n_\text{kd(kn)}$ the noise photons in excess of the quantum limit added by the amplifier in each mode of operation. $G_k$ is the KIPA gain, equivalent to the amplification level $G_A$ when in degenerate mode.

In Fig.~\ref{fig4}b we plot the input-referred  number of photons recorded in the presence of an applied coherent tone, with the KIPA in three different configurations: degenerate mode, non-degenerate mode and off. Both the degenerate and non-degenerate measurements are taken at a pump power of $P_\text{pump} = -19.3\,\text{dBm}$, which provides gains of $31\,\text{dB}$ and $24.5\,\text{dB}$, respectively. We observe an 18~dB enhancement in the SNR when operating in non-degenerate mode relative to using the HEMT as the first-stage amplifier (KIPA off) and an additional 6~dB of SNR enhancement when operating in degenerate mode (see Fig.~\ref{fig4}b), which is predicted by Eqs.~\ref{eqn:ntd}-\ref{eqn:ntnd} in the limit of large gain ($G_k \to \infty$). 

In order to extract the noise parameters we measure $n_\text{td}$ and $n_\text{tn}$ as a function of $G_k$ in the absence of an applied coherent tone and plot the results in Fig.~\ref{fig4}c. Fits to the curves yield $n_\text{td}^\infty = n_\text{th}/2 + n_{kd}/2 + 1/4 = 0.31(5)$, $n_\text{tn}^\infty = 2n_\text{th} + n_{kn} + 1 = 1.18(9)$ and $n_\text{sys} = 80.0(46)$. $n_\text{kd}$ and $n_\text{kn}$ result from internal losses in the amplifier that mix bath photons into the output the field, whilst $n_\text{th} = 1/[\exp{(\hbar\omega_0/k_BT_\text{em})}-1]$ is given by the Bose-Einstein distribution, with $T_\text{em}$ the temperature of the electromagnetic field. It is important to note that $T_\text{em}$ can be a different value to the physical temperature of the refrigerator $T_f$ due to insufficient filtering and attenuation of the measurement lines. In our noise measurement we are not able to separate the individual contributions of $n_\text{kd}/n_\text{kn}$ and $n_\text{th}$. At the maximum degenerate gain measured in Fig.~\ref{fig4}c (35~dB) we find an input-referred noise of 0.32(3) photons, a value close to the theoretical quantum limit of 0.25 photons.

\section{Conclusion}
We have presented a simple and versatile microwave parametric amplifier called a KIPA, fabricated from a thin film of NbTiN. For the parameter space tested, we report above $40\,\text{dB}$ of phase insensitive gain and up to $50\,\text{dB}$ of phase sensitive gain. Our device features an exceptionally high input $1\,\text{dB}$-compression point of $-69.5\,\text{dBm}$ for $20\,\text{dB}$ of gain, making it suitable for a wide range of cryogenic microwave measurements. Using input-output theory we have been able to model our device as an ideal DPA, with excellent agreement between theory and experiment. The close DPA approximation makes the KIPA an extremely effective microwave squeezer, which we demonstrate through mapping its squeezing transformation out to $G_A = 30\,\text{dB}$. We find the amplifier operates close to the quantum noise limit with an input-referred noise of 0.32(3) photons in phase sensitive mode.

Future experiments will focus on exploring the noise squeezing properties of the KIPA by using a second KIPA as a following amplifier \cite{castellanos-beltran_2008,bienfait_2017}. The large levels of deamplification without distortion observed here sets an upper bound to its noise squeezing capabilities as any loss present will act to mix in the vacuum state and reduce squeezing (see Supplemental Materials for a theoretical analysis of squeezing in the presence of loss). However, we see no observable signs of loss in the current device and note that planar NbTiN resonators can reach exceptionally large internal quality factors ($Q_i > 10^6$ \cite{barends_2010,bruno_2015}), making this an attractive system to perform noise squeezing.

Squeezing levels in excess of $17$~dB would surpass the fault-tolerant threshold for measurement-based quantum computing with cluster-states \cite{menicucci_2014}. Our estimates of the internal quality factor suggests that squeezing may already be above this threshold in the current device. Combining this with the already demonstrated ability to engineer non-Gaussian states of light in superconducting circuits \cite{hofheinz_2008,grimm_2020} raises the exciting possibility of utilising the KIPA as a resource in continuous variable quantum computing. Finally, high kinetic inductance NbTiN resonators can display excellent magnetic field compatibility, with $Q_i > 10^5$ at fields up to 6~T reported \cite{samkharadze_2016}, we therefore envisage utilising this amplifier in applications such as electron spin resonance spectroscopy, where the KIPA can serve as both the microwave cavity and first-stage amplifier to push the boundary of spin detection sensitivity \cite{bienfait_2016,eichler_2017}.

\begin{acknowledgments}
J.J.P. is supported by an Australian Research Council Discovery Early Career Research Award (DE190101397). J.J.P. and A.M. acknowledge support from the Australian Research Council Discovery Program (DP210103769). A.L.G. is supported by the Australian Research Council, through the Centre of Excellence for Engineered Quantum Systems (CE170100009) and Discovery Early Career Research Award (DE190100380). A.M. is supported by the Australian Department of Industry, Innovation and Science (Grant No. AUSMURI000002). W.V. acknowledges financial support from Sydney Quantum Academy, Sydney, NSW, Australia. The authors acknowledge support from the NSW Node of the Australian National Fabrication Facility. We thank Robin Cantor and STAR Cryoelectronics for sputtering the NbTiN film. The authors thank Nicolas Menicucci and Patrice Bertet for helpful discussions.
\end{acknowledgments}

\bibliography{kipa_MS}
\bibliographystyle{unsrtnat}

\end{document}


\title{Supplemental Material: A near-ideal degenerate parametric amplifier}

\author{Daniel J. Parker}
\affiliation{School of Electrical Engineering and Telecommunications, UNSW Sydney, Sydney, NSW 2052, Australia}
\author{Mykhailo Savytskyi}
\affiliation{School of Electrical Engineering and Telecommunications, UNSW Sydney, Sydney, NSW 2052, Australia}
\author{Wyatt Vine}
\affiliation{School of Electrical Engineering and Telecommunications, UNSW Sydney, Sydney, NSW 2052, Australia}
\author{Arne Laucht}
\affiliation{School of Electrical Engineering and Telecommunications, UNSW Sydney, Sydney, NSW 2052, Australia}
\author{Timothy Duty}
\affiliation{School of Physics, UNSW Sydney, Sydney, NSW 2052, Australia}
\author{Andrea Morello}
\affiliation{School of Electrical Engineering and Telecommunications, UNSW Sydney, Sydney, NSW 2052, Australia}
\author{Arne L. Grimsmo}
\affiliation{Centre for Engineered Quantum Systems, School of Physics, The University of Sydney, Sydney, Australia}
\author{Jarryd J. Pla}
\affiliation{School of Electrical Engineering and Telecommunications, UNSW Sydney, Sydney, NSW 2052, Australia}

\date{\today}

\maketitle

\tableofcontents

\section{Device Fabrication}
The KIPA is fabricated on a $350\,\mu\text{m}$ thick high-resistivity silicon wafer. The wafer is cleaned with a piranha solution (a mixture of sulfuric acid, water and hydrogen peroxide) followed by an HF etch of the natural silicon dioxide before having a $9.5\,\text{nm}$ thick film of NbTiN sputtered on the surface (STAR Cryoelectronics). To define the pattern we perform a standard electron beam lithography process using AR-P 6200 (9\%) as a positive resist. Reactive Ion Etching (RIE) with $\text{CF}_4$ and Ar is used to etch the NbTiN in the exposed regions of the chip. After the RIE step any residual resist mask is removed using solvents before the device is bonded to a printed circuit board and measured. 

\section{Experimental Setup}
All measurements are performed with the device situated at the mixing chamber plate ($T\sim20\,\text{mK}$) of a dilution refrigerator.

\subsection{Wiring}

\textbf{The Pump Line:}
A microwave source (E8267D, Keysight Technologies) supplies the pump tone for all experiments via a high pass filter (HFCN-9700+, Mini-Circuits) used to reduce microwave source subharmonics. A $10\,\text{dB}$ cryogenic attenuator is used at the $4\,\text{K}$ temperature stage, followed by two $3\,\text{dB}$ attenuators at the $900\,\text{mK}$ and $100\,\text{mK}$ stages, respectively. The pump line connects to the KIPA via a diplexer (DPX-1114, Marki Microwave) at the $20\,\text{mK}$ stage (shown in Fig.~1a of the main text), which provides $> 40$~dB of rejection at the signal frequency $\omega_0/2\pi = 7.2$~GHz.\\

\textbf{The Signal Line:}
Three $20\,\text{dB}$ attenuators are used to minimise the transmission of thermal noise to the device, and are situated at the $4\,\text{K}$, $900\,\text{mK}$ and $20\,\text{mK}$ stages, respectively. The signal line then connects to the RF port of the bias-T (PE1615, Pasternack Enterprises), as shown in Fig.~1a (main text).\\

\textbf{The DC Line:}
The DC line connects to the bias-T via a 1~dB attenuator at 4~K and two low pass filters at 4~K and 100~mK (VLF-7200+ and VLF-105+, Mini-Circuits Technologies), blocking room-temperature noise at pump and signal frequencies. The DC line breaks out to a copper wire that is thermalised to a bobbin fixed to the 20 mK plate before connecting to the DC port of a bias tee (shown in Fig.~1a).\\

\textbf{The Detection Path:}
A cryogenic circulator (Quinstar Technology, CTH0508KCS) routes the reflected output of the KIPA through the detection chain (shown in Fig.~1a). A high-rejection bandpass filter (Micro-Tronics Inc, BPC50403-01) immediately follows and attenuates any power at the pump frequency that may leak through the diplexer. A double isolator (Quinstar Technology, CTH0508KCS $\times 2$) at $20\,\text{mK}$ connects the output of the bandpass filter to a cryogenic HEMT low noise amplifier (Low Noise Factory, LNF-LNC0.3\_14A) situated at $4\,\text{K}$.

\subsection{VNA Measurements}
Port 1 of a vector network analyzer (Rohde \& Schwarz, ZVB-20) is connected to the signal line via an attenuator, used to reduce the minimum signal power of the network analyzer. We use a low noise amplifier (Mini-Circuits, ZX60-06183LN+) at the output of the detection chain, which connects to port 2 of the VNA.

The data presented in Fig.~1c (main text) was collected with the pump source disabled. We apply a DC voltage (Yokogawa Electric, GS2000) to the DC line, in series with a $\sim 10\,\text{k}\Omega$ resistor at room temperature. We observe a $2\pi$ phase shift in the frequency response measured with the VNA, as expected for a $\lambda/4$ resonator measured in reflection in the over-coupled regime (see \cref{sec:gain_features}). A linear fit to the first $100\,\text{MHz}$ of the phase response is used to estimate the line-delay of our setup and is subtracted from the complete phase response. The phase is then increased by $\pi$ to correct for the expected phase offset that is removed by the fit to the line-delay. To model the resonance frequency shift, we fit a quadratic polynomial to the resonance frequency as a function of the square of the current.

The $14.318\,\text{GHz}$ pump is then enabled with a $0.834\,\text{mA}$ DC bias current for the measurement of the phase insensitive gain in Fig.~2a (main text). We use the VNA to probe the magnitude response about half the pump frequency. To estimate the baseline of the magnitude response, we disable the pump but leave the bias current active, which yields an approximately flat magnitude response (see \cref{sec:losses}). We subtract the magnitude response of the baseline measurement from the magnitude response of the gain curve to obtain the data presented in Fig.~2a (main text). For an detailed explanation of the fitting procedure, refer to \cref{sec:phase_insens_gain}.

To study the phase sensitive gain, we operate the ZVB-20 as a spectrum analyzer, using it to measure the incident power on Port 2. The signal line is connected to another E8267D microwave source (Keysight Technologies) via an attenuator and is configured for linear phase modulation at half the pump frequency ($7.1905\,\text{GHz}$) and $\sim -112\,\text{dBm}$ of signal power at the sample. The pump and signal sources are phase locked using a $1\,\text{GHz}$ reference clock. With the VNA configured for a zero-span measurement and triggered off the edge of each phase ramp, we obtain the data presented in Fig.~2b of the main text. Again, we disable the pump, measure the baseline and subtract the mean reflected baseline power from each measurement to obtain the phase sensitive gain. Due to slow phase drifts between the VNA local oscillator and the signal tone, we repeat each measurement 40 times, and use the maximum of the cross correlation between pairs of traces to align the data before averaging. We repeat this measurement for a range of signal powers and pump powers, and use the maximum of the gain curve to define the degenerate gain, as presented in Fig.~2c. The compression power is determined by the signal power where the pre-saturation gain drops by $1\,\text{dB}$. We define the pre-saturation gain by the average of the degenerate gains measured for the 10 smallest signal powers.

\subsection{Coherent State Measurement}
For the remainder of the measurements, the output of the detection chain is connected to a homodyne detection setup consisting of an IQ mixer (Marki Microwave, IQ4509), with the local oscillator supplied by another independent ultra low phase noise microwave source (Keysight Technologies, E8267D) which is phase locked with a $1\,\text{GHz}$ reference clock to the pump and signal sources. The local oscillator frequency is set to $7.1905\,\text{GHz}$. The I and Q outputs of the mixer connect to $1.9\,\text{MHz}$ low pass filters (Mini-Circuits, SLP-1.9+) followed by two $5~\times$ pre-amplifiers (Stanford Research
Systems, SIM914) connected in series. I and Q are then digitized using a data acquisition card (Keysight Technologies, M3300A) configured with a sample rate of $6.25\,\text{MHz}$.

The ellipse measurements (depicted in Fig.~3b of the main text) were performed with the pump and local oscillator phases fixed, while the signal phase is stepped. Each (I, Q) pair is measured by averaging $10^6$ samples collected at each phase. The entire phase sweep is performed in less than $60\,\text{s}$ to minimise errors due to slow phase drift between the signal and pump. Before each measurement, we calibrate the phase of the local oscillator by rotating the IQ-plane in software to measure the angle that produces the least variance on Q. We refine phase calibration by repeating the procedure three times in order to ensure measurement consistency despite small channel imbalances between I and Q. We measure 16 repetitions of the phase sweep and software rotate each dataset to further minimise the variance on Q due to slow phase drifts in the setup. The repetitions are aligned by maximising the pairwise cross-correlation of I (Q) as a function of the signal phase, and then averaged to produce the data presented in Fig.~3b.

We interleave a measurement with the pump off to measure the circular response of the reflected signal in the IQ-plane. We measure $G_S$ ($G_A$) by taking the ratio of the peak to peak amplitudes of the pump off response and pump on response for Q (I) after averaging. These results are plotted in Fig.~3c (main text). Phase calibration is performed with the pump on and the calibration phase is kept after the pump is disabled.

\subsection{Noise Measurements}
To collect the data presented in Fig.~4 of the main text, we replace the room temperature amplifier with a low noise HEMT amplifier (Low Noise Factory, LNF-LNR1\_15A) for improved noise performance. For the measurement in Fig.~4c the signal source is disabled and the input to the KIPA is defined by the noise produced by the nearest microwave attenuator; i.e. approximately $1/2$ a photon at $7.1905\,\text{GHz}$ and $20\,\text{mK}$.

\section{Hamiltonian of a Kinetic Inductance Parametric Amplifier}\label{sec:Ham}
\subsection{Zero Bias}
\begin{figure}
\centering
\includegraphics[scale=0.65]{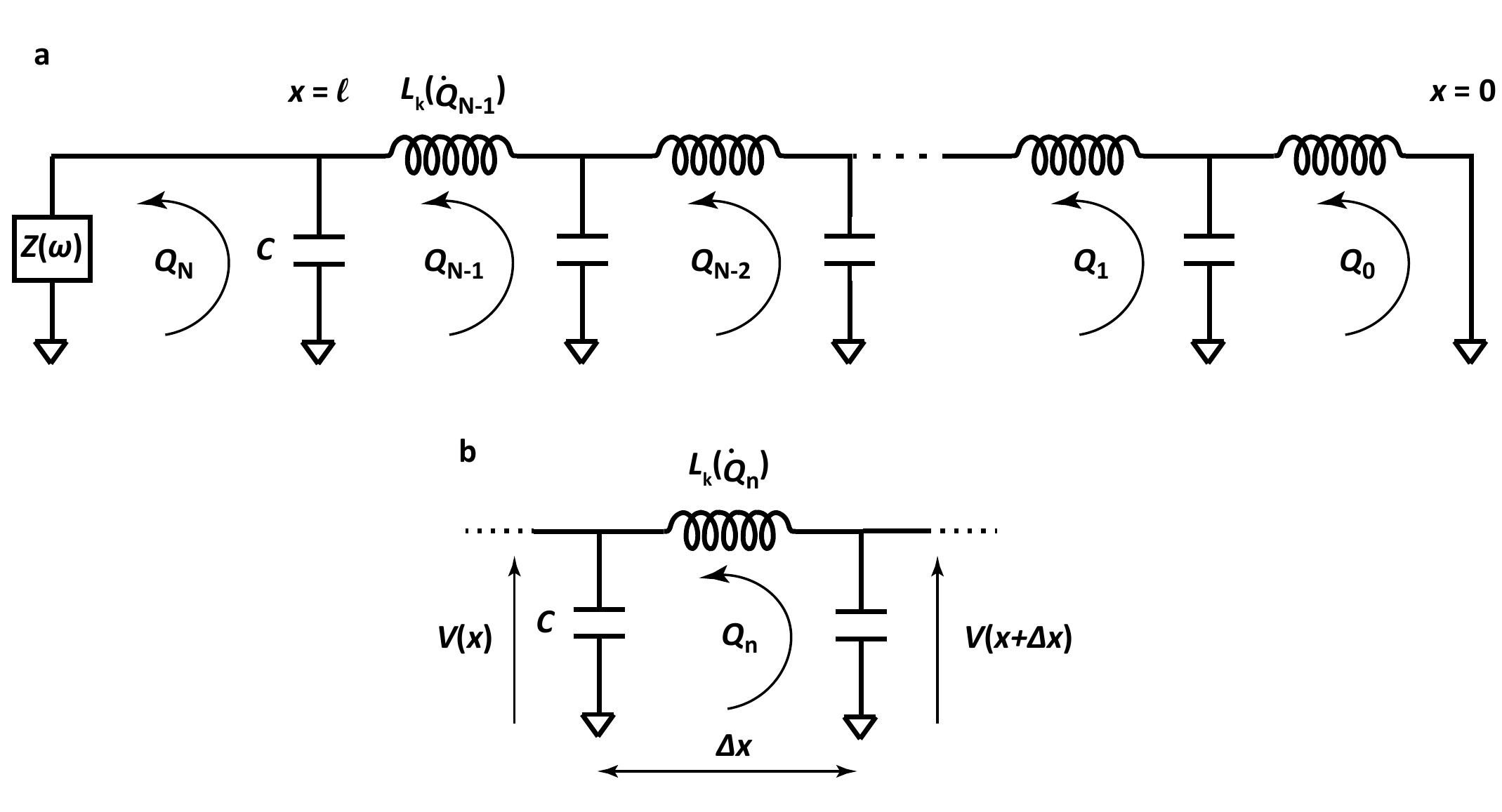}
\caption{(a) Telegrapher's model of a kinetic inductance resonator. (b) A single loop of the Telegrapher's model.\label{fig:kir}}
\end{figure}

A kinetic inductance can be described as a nonlinear inductance
\begin{equation}\label{eq:LI}
    L_k(I) = L_0\left(1 + \frac{I^2}{I_*^2}\right).
\end{equation}
We consider a `telegrapher model' for a kinetic inductance resonator, as illustrated in Fig.~S\ref{fig:kir}a and wish to write down the Lagrangian for this system. The circuit has a capacitance $C$ and kinetic inductance $L_k(I)$ per unit length. We assume that the kinetic inductance is far greater than the geometric inductance ($L_g$) along the transmission line (i.e. $L_0 \gg L_g$). Because the inductance depends on current, and current is related to charge in a straight-forward manner, we find it convenient in this situation to formulate the Lagrangian with charge as the coordinate. We therefore use the `loop charge' approach described in Ref.~\cite{jascha_2016}. The loop charges are related to the current across the inductors and charge on the capacitors through:
\begin{align}
I_n ={}& \dot Q_n,\\
q_n ={}& Q_{n} - Q_{n-1},
\end{align}
respectively. Note that in this particular geometry, the loop charge equals the cumulative charge $Q_n = \sum_{k=0}^{n-1} q_k$.

Applying Kirchoff's voltage law around a single loop in the Telegrapher circuit model (see Fig.~S\ref{fig:kir}b):
\begin{equation}
    \begin{aligned}
        V(x+\Delta x,t) ={}& V(x,t) + L_k(I_n)\Delta x\dot I_n,\\
        \frac{V(x+\Delta x,t)-V(x,t)}{\Delta x} ={}& L_k(I_n)\dot I_n,\\
         \to \partial_x V(x,t) ={}& L_k(I)\partial_t I,
    \end{aligned}
\end{equation}
where in the third line we take a continuum limit $\Delta x \to 0$. This is the well-known Telegrapher's equation, with a nonlinear inductance as provided in \cref{eq:LI}, and describes the relevant equation of motion for our circuit. The Telegrapher equation can equivalently be expressed as:
\begin{equation}\label{eq:EOM}
    \frac{1}{C}\partial_x^2 Q(x,t) = L_k(\partial_t Q)\partial_t^2 Q,
\end{equation}
with $I = \partial_t Q$. The Lagrangian for the system that reproduces the Telegrapher's equation with its Euler-Lagrange equation is found to be:
\begin{equation}\label{eq:Lkipa:unbiased}
    \begin{aligned}
        \mathcal L_\kir
        ={}& \half \sum_{n=0}^{N-1} \left[L_0\Delta x\left(1+\frac{1}{6}\frac{\dot Q_n^2}{I_*^2}\right) \dot Q_n^2 - \frac{1}{C \Delta x} (Q_{n+1}-Q_{n})^2\right],\\
        \to{}& \half \int_0^l \dd x \left[ L_0\left(1+\frac{1}{6}\frac{(\partial_t Q)^2}{I_*^2}\right)(\partial_t Q)^2 - \frac{1}{C} (\partial_x Q)^2\right],\\
        ={}& \half \int_0^l \dd x \left[ L_0 (\partial_t Q)^2 - \frac{1}{C} (\partial_x Q)^2\right]
        + \frac{L_0}{12I_*^2} \int_0^l \dd x (\partial_t Q)^4,
    \end{aligned}
\end{equation}
where in the second line we once again take a continuum limit $\Delta x \to 0$. We note that this form of the Lagrangian differs from the work of \citeauthor{yurke_2006}, but correctly reproduces the classical Telegrapher's equations for a kinetic inductance transmission line (\cref{eq:EOM}) assumed in recent work \cite{erickson_2017,malnou_2021}.

The canonical momentum corresponding to $Q$ is:
\begin{equation}
    \Phi = \frac{\partial \mathcal L_\kir}{\partial \dot Q} = L_0 \partial_t Q + \frac{L_0}{3I_*^2} (\partial_t Q)^3,
\end{equation}
and the Hamiltonian is given by:
\begin{equation}
    \begin{aligned}
    H ={}& \int_0^l \dd x \Phi \partial_t Q - \mathcal L\\
    ={}& \half \int_0^l \dd x \left[ L_0 (\partial_t Q)^2 + \frac{1}{C} (\partial_x Q)^2\right]
    + \frac{3L_0}{12I_*^2} \int_0^l \dd x (\partial_t Q)^4.
    \end{aligned}
\end{equation}

To express this in terms of $\Phi$ and $Q$ we use the approximation:
\begin{equation}
    \partial_t Q = \frac{1}{L_0} \Phi - \frac{1}{3I_*^2}(\partial_t Q)^3
    = \frac{1}{L_0} \Phi - \frac{1}{3I_*^2L_0^3} \Phi^3
    + \mathcal O\left(\frac{1}{I_*^4}\right),
\end{equation}
and:
\begin{subequations}\label{eq:dQ24}
    \begin{align}
        (\partial_t Q)^2 \simeq{}& \frac{1}{L_0^2} \Phi^2 - \frac{2}{3I_*^2L_0^4} \Phi^4,\\
        (\partial_t Q)^4 \simeq{}& \frac{1}{L_0^4} \Phi^4.
    \end{align}
\end{subequations}

Thus, to first order in $1/I_*^2$ we find:
\begin{subequations}\label{eq:Hkipa:unbiased}
    \begin{align}
    H_\kir ={}& H_0 + H_1,\\
    H_0 ={}& \half \int_0^l \dd x \left[ \frac{1}{L_0} \Phi^2 + \frac{1}{C} (\partial_x Q)^2\right],\\
    H_1 ={}& -\frac{1}{12I_*^2 L_0^3} \int_0^l \dd x \Phi^4.
    \end{align}
\end{subequations}

\subsubsection{Mode Expansion for the $\lambda/4$ Resonator}

We start by finding the modefunctions of a linear ($I_* \to \infty$) $\lambda/4$ resonator. In this case, the Euler-Lagrange equation corresponding to $\mathcal L_\kir$ is the Telegraper's equation:
\begin{equation}\label{eq:wave_equation}
    v_0^2 \partial_x^2 Q = \partial_t^2 Q,
\end{equation}
with $v_0 = 1/\sqrt{L_0 C}$ the linear phase velocity.

The $\lambda/4$ resonator is shorted at $x=0$ corresponding to a boundary condition of zero voltage, or $(Q_1 - Q_0)/C\Delta x \to \partial_x Q(x=0)/C = 0$ in the continuum limit. At $x=l$ we leave the boundary condition general by taking an impedance $Z(\omega)$ to ground and imposing Ohm's law $V(x=l) = -Z(\omega)I(x=l)$ (for an $I$ convention defined in Fig.~\ref{fig:kir}) at the boundary, with $V(x=l) = \partial_x Q(x=l)/C$ and $I(x=l) = \partial_t Q(x=l)$. In summary:
\begin{subequations}\label{eq:bcs}
    \begin{align}
        \partial_x Q(x=0) ={}& 0 \qquad \qquad \qquad \qquad \quad \quad (\text{short circuit}),\\
        \partial_t Q(x=l) ={}& -\frac{1}{Z(\omega) C} \partial_x Q(x=l) \quad (\text{impedance $Z(\omega)$ to ground}).
    \end{align}
\end{subequations}
An open (short) boundary condition at $x=l$ is recovered in the limit $Z(\omega) \to i\infty$ ($Z(\omega) \to i0$).

We use an ansatz:
\begin{equation}
    Q(x,t) =i \sum_m A_m \cos\left( k_m x + \phi_m \right) [a_m\dg(t) - a_m(t)],
\end{equation}
with $a(t) = a e^{-i\omega_m t}$ and $k_m = \omega_m/v_0$. The first boundary condition is met by setting $\phi_m = 0$. The second boundary condition gives:
\begin{equation}
\tan(k_m l) = \frac{\partial_t Q_m(t)}{\omega_m Q_m(t)}\frac{Z(\omega_m)}{Z_0},
\end{equation}
where $Z_0 = \sqrt{L_0/C}$ is the characteristic impedance of the $\lambda/4$ resonator and $Q_m(t)$ is the time-dependent component of $Q(x,t)$ oscillating at $\omega_m$. The equation must in general be solved numerically for $k_m$. In the case of an open where $Z(\omega)\to i\infty$, we simply have:
\begin{subequations}\label{eq:km_open}
\begin{align}
\cos(k_m l) ={}& 0 \Rightarrow k_m = \frac{(2m+1)\pi}{2l} \qquad (\text{open}),\\
\omega_m ={}& \frac{(2m+1)\pi v_0}{2l}.
\end{align}
\end{subequations}

The band-stop filter presents the resonator with a large impedance for frequencies within the stop band. To simplify the following analysis we assume an infinite impedance, i.e. an open boundary condition at $x = l$, which allows us to utilize the relations in~\cref{eq:km_open}. 

In the linear case, the canonical momentum is just $\Phi = L_0 \dot Q$, i.e.:
\begin{equation}
    \Phi(x,t) = -\sum_{m=0}^\infty L_0 \omega_m A_m \cos\left( k_m x \right) [a_m\dg(t) + a_m(t)].
\end{equation}
Quantization proceeds by imposing the commutation relations $[a_m(t),a_n\dg(t)]=\delta_{nm}$. The normalization constants $A_m$ are determined by inserting $Q(x,0)$ and $\Phi(x,0)$ into $H_\kir$ and requiring:
\begin{equation}\label{eq:Hlin}
    H_0 = \sum_{m=0}^\infty \hbar\omega_m\left(a_m\dg a_m + \frac12\right),
\end{equation}
which leads to $A_m = 1/\sqrt{lL_0\omega_m}$. We therefore find:
\begin{subequations}\label{eq:fields}
\begin{align}
Q(x,t) ={}& i \sum_{m=0}^\infty \sqrt{\frac{\hbar}{L_T \omega_m}} \cos\left( k_m x \right) [a_m\dg(t) - a_m(t)],\\
\Phi(x,t) ={}& -\frac{1}{l}\sum_{m=0}^\infty \sqrt{\hbar L_T\omega_m} \cos\left(k_m x \right) [a_m\dg(t) + a_m(t)],
\end{align}
\end{subequations}
where we define $L_T = L_0 l$ as the total zero-bias kinetic inductance of the resonator.

More generally, we can interpret~\cref{eq:fields} as a change of variables from $\{Q(x,t), \Phi(x,t)\}$ to $\{a_m(t), a_m\dg(t)\}$, subject to the spatial boundary constraints. Inserting the form of $\Phi(x,t)$ into~\cref{eq:Hkipa:unbiased}c, keeping only the fundamental mode and dropping fast rotating terms and constants, we find:
\begin{equation}
    \begin{aligned}
	    H_1 ={}& -\frac{(\hbar\omega_0)^2}{32L_TI_*^2}(a\dg + a)^4,\\
	    \approx{}& -\frac{3(\hbar\omega_0)^2}{16L_TI_*^2}(2a\dg a + a{\dg}^2a^2),\\
	    ={}& \hbar K a\dg a + \frac{\hbar K}{2}a{\dg}^2a^2.
    \end{aligned}
\end{equation}

The Kerr nonlinearity is thus:
\begin{equation}
    K = -\frac{3}{8}\frac{\hbar\omega_0}{L_TI_*^2}\omega_0.
\end{equation}
Here $E_* \equiv L_TI_*^2/2$ can be interpreted as a characteristic energy stored in an inductor with inductance $L_T$ and current $I_*$, which is also related to the superconducting pairing energy $E_p = L_TI_*^2$ \cite{zmuidzinas_2012}.

\subsection{Current Bias}

\subsubsection{Mode Expansion}

In the presence of a current bias, we modify the boundary condition at $x=l$ to be:
\begin{equation}
\partial_t Q(x=l) = -\frac{1}{Z(\omega) C} \partial_x Q(x=l) + I_b.
\end{equation}
We take the impedance to be a stop-band filter at the relevant resonator mode frequencies:
\begin{equation}
    Z_s(\omega)=
    \left\{
    \begin{array}{ll}
    i\infty & \omega \in \Omega_0 \\
    50~\Omega & \omega \in \Omega_1
    \end{array}
    \right.,
\end{equation}
where $\Omega_0$ represents the frequency band over which we have standing resonator modes, and $\Omega_1$ covers the impedance matched frequency band, where we will have traveling waves. For $I_b=0$ we can then write:
\begin{equation}
    \begin{aligned}
    Q(x,t) ={}& i \sum_m \sqrt{\frac{\hbar}{L_T\omega_m}} \cos\left( k_m x \right) [a_m\dg(t) - a_m(t)]\\
    &+ i \int_{\Omega_1} \dd\omega \sqrt{\frac{\hbar}{\pi\omega v_0 L_0}}\cos\left(\frac{\omega x}{v_0}\right)[b_\omega\dg(t) - b_\omega(t)],
    \end{aligned}
\end{equation}
with $b_\omega\dg(t) = b_\omega\dg e^{i\omega t}$. We already know that the first term satisfies the boundary conditions at frequencies $\omega_m$ from our previous analysis, with $\omega_m,k_m$ given in~\cref{eq:km_open}. For frequencies $\omega\in \Omega_1$, on the other hand, the circuit is modeled as a semi-infinite matched transmission line connected to ground at $x=0$.

In the presence of a current bias, we simply add to $Q(x,t)$ a term $q_b(x,t)$ where $\dot q_b(x,t) = I_b(x,t)$. We assume that the pump frequencies are in the traveling wave band $\Omega_1$. Equivalently, we can replace $b_\omega \to b_\omega + \beta(\omega)$, with $\beta(\omega)$ the frequency component of $q_b(t)$ at frequency $\omega$. This can be interpreted as separating the continuum mode into a strong average coherent component $\beta(\omega)$ and a fluctuation (or quantum) term $b_\omega$. We take the pump to be infinitely narrow in frequency, and therefore set $\beta(\omega) = \beta_p \delta(\omega-\omega_p) + \beta_\dc\delta(\omega)$. The DC component is independent of space and trivially satisfies both the wave equation~\cref{eq:wave_equation} and the short $\partial_x Q = 0$ boundary condition. We therefore finally have:
\begin{equation}
    \begin{aligned}
    Q(x,t) ={}& i \sum_m \sqrt{\frac{\hbar}{L_T\omega_m}} \cos\left( k_m x \right) [a_m\dg(t) - a_m(t)]\\
    &+ i \int_{\Omega_1} \dd\omega \sqrt{\frac{\hbar}{\pi\omega v_0 L_0}}\cos\left(\frac{\omega x}{v_0}\right)[b_\omega\dg(t) - b_\omega(t)] + q_p(x,t) + q_\dc(t),
    \end{aligned}
\end{equation}
where:
\begin{subequations}
    \begin{align}
        \dot q_\dc(t) ={}& I_\dc,\\
        \dot q_p(t) ={}& I_p(x,t) = \cos\left(\frac{\omega_p x}{v_0}+\phi_p\right) I_p(t).
    \end{align}
\end{subequations}

In our experiments we apply a pump tone $I_p(x,t)$ that oscillates at a frequency very close to $2\omega_0$ and we will therefore assume $\omega_p = 2\omega_0$ for simplicity. Furthermore, the boundary condition in \cref{eq:bcs}a implies that $\phi_p=0$. Thus:
\begin{equation}
    I_p(x,t) = \cos(2k_0x) I_p(t).
\end{equation}

As before, we will use the strategy of taking the mode expansion of $\Phi(x,t)$ in the absence of any nonlinearity, and substitute this back into the nonlinear Hamiltonian $H_1$. Using $\Phi(x,t) = L_0 \dot Q(x,t)$ we find:
\begin{equation}
    \begin{aligned}
    \Phi(x,t) ={}& -\frac{1}{l} \sum_m \sqrt{\hbar L_T\omega_m} \cos\left( k_m x \right) [a_m\dg + a_m]\\
    & - \frac{1}{v_0}\int_{\Omega_1} \dd\omega \sqrt{\frac{\hbar \omega Z_0}{\pi}}\cos\left(\frac{\omega x}{v_0}\right)[b_\omega\dg + b_\omega] + L_0 \cos(2k_0x) I_p(t) + L_0 I_\dc.
    \end{aligned}
\end{equation}

\subsubsection{Hamiltonian}

For the linear Hamiltonian we have as before:
\begin{equation}
    H_0 = \sum_{m=0}^\infty \hbar\omega_m\left(a_m\dg a_m + \frac12\right),
\end{equation}
where we have kept only the resonator modes and dropped the continuum modes in $\Omega_1$. To proceed we will substitute the (Schr\"odinger picture) flux field into $H_1$. In general this will lead to coupling (e.g. cross-Kerr) between resonator and quantum continuum modes. However, given that the nonlinearity of the KIPA is extremely weak, we will neglect the quantum fluctuations of the current, i.e. drop the $b_\omega$ modes completely from the nonlinearity. We also, for simplicity, truncate to the fundamental resonator mode, and thus use:
\begin{equation}
    \begin{aligned}
    \Phi(x,t) ={}& -\frac{1}{l} \sqrt{\hbar L_T\omega_0} \cos\left( k_0 x \right) (a\dg + a)
    + L_0 \cos(2k_0x) I_p(t) + L_0 I_\dc.
    \end{aligned}
\end{equation}

Dropping fast rotating terms in $a$, $a\dg$ from $H_1$:
\begin{equation}
    \begin{aligned}
    H_1 ={}& - \frac{3 (\hbar\omega_0)^2}{16 I_*^2 L_T}\left( 2 a\dg a + a{\dg}^2 a^2\right)\\
    &- \frac{\hbar \omega_0}{8I_*^2}
    \left(2I_\dc^2 + 2I_\dc I_p(t) + I_p^2(t)\right) \left( 2 a\dg a + a{\dg}^2 + a^2\right).
    \end{aligned}
\end{equation}

We take the time-dependent AC current amplitude to be:
\begin{equation}
    I_p(t) = \frac{I_p}{2}\left(e^{-i(\omega_p t + \varphi_p)} + e^{i(\omega_p t + \varphi_p)}\right).
\end{equation}

Substituting $I_p(t)$ into $H_1$ and transferring to a frame rotating at $\omega_p/2$, the KIPA Hamiltonian becomes:
\begin{equation}\label{eq:HKIPArot}
    \begin{aligned}
	    H_\kir ={}& \hbar\left(\omega_0 + \delta_\dc + \delta_p + K -\frac{\omega_p}{2}\right) a\dg a + \frac{\hbar\xi}{2}a{\dg}^2 + \frac{\hbar\xi^*}{2}a^2 + \frac{\hbar K}{2}a{\dg}^2 a^2,
    \end{aligned}
\end{equation}
where any fast rotating pump terms have been ignored. We thus define the following important Hamiltonian parameters:
\begin{subequations}\label{eq:Hamparam}
    \begin{align}
        \delta_\dc ={}& -\frac{1}{2}\frac{I_\dc^2}{I_*^2}\omega_0,\\
        \delta_p ={}& -\frac{1}{8}\frac{I_p^2}{I_*^2}\omega_0,\\
        K ={}& - \frac{3}{8}\frac{\hbar\omega_0}{L_TI_*^2}\omega_0,\\
        \xi ={}& -\frac{1}{4}\frac{I_\dc I_p}{I_*^2}\omega_0e^{-i\varphi_p}.
    \end{align}
\end{subequations}

We note that the term $\delta_p$ arises from the square of the pump current, which has a non-zero average value of $I_p^2/2$ and therefore causes an effective detuning of the cavity frequency.

As a sanity check, in the absence of a pump tone (i.e. $I_p = 0$) we find the resonance frequency of the cavity to be:
\begin{equation}
    \omega_0^\prime \approx \omega_0\left(1-\frac{1}{2}\frac{I_\dc^2}{I_*^2}\right),
\end{equation}
neglecting the Kerr term. For $(I_\dc/I_*)^2 \ll 1$, which is the approximation used in~\cref{eq:dQ24}, we can write:
\begin{equation}
    \omega_0^\prime \approx \frac{\omega_0}{\sqrt{1+I_\dc^2/I_*^2}} = \frac{\pi}{2l\sqrt{CL_k(I_\dc)}},
\end{equation}
which is the fundamental frequency of a $\lambda/4$ resonator with a per unit length capacitance of $C$ and inductance $L_k(I_\dc)$ (provided by~\cref{eq:LI}), as expected.

\subsection{DPA Approximation}
A comparison of the expressions for $K$ (\cref{eq:Hamparam}c) and $\xi$ (\cref{eq:Hamparam}d) reveals why the KIPA functions as an ideal DPA: the photon energy is a minuscule fraction of the characteristic nonlinear inductive energy (i.e. $\hbar\omega_0/(L_T I_*^2) \ll 1$) by virtue of $I_*$ being large. 

We estimate the per-square kinetic inductance $L_\square$ of the NbTiN film by performing a simulation of the full KIPA device structure (using the software Sonnet) and adjusting $L_\square$ until we obtain the measured zero current bias resonance frequency $\omega_0$. We find $L_T = L_\square(l/w) = 3.84$~nH (with $l$ and $w$ the length and width of the $\lambda/4$ resonator), which together with the measured value of $I_*$ gives $K \approx 0.13$ Hz, a completely negligible quantity relative to all other system parameters. Compared to the coupling rate, we achieve the ratio: $\kappa/K > 10^{8}$, greater than the typical values of $\kappa/K < 10^{4}$ seen in JPAs \cite{boutin_2017}. Because the Kerr term is so small, we approximate the Hamiltonian for the KIPA with the Hamiltonian for the ideal DPA for the remainder of this work:
\begin{equation}\label{eq:kipadpa}
    \begin{aligned}
	    H_\kir \approx{}&  H_\text{DPA} = \hbar\Delta a\dg a + \frac{\hbar\xi}{2}a{\dg}^2 + \frac{\hbar\xi^*}{2}a^2
    \end{aligned}
\end{equation}
with:
\begin{equation}
    \Delta = \omega_0 + \delta_\dc + \delta_p - \omega_p/2
\end{equation}
which is identical to~\cref{eq:HKIPArot} with the Kerr terms neglected.

\section{Input-Output Theory for a DPA}\label{sec:IO}
\begin{figure}
	\centering
	\includegraphics[width=.7\linewidth]{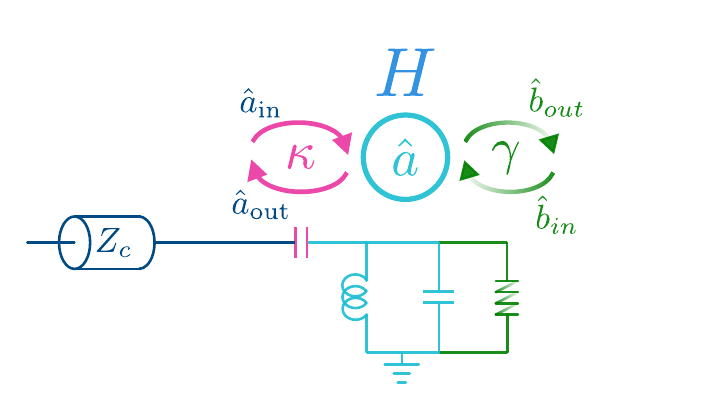}
	\caption{\label{fig:io_model}
		The single port input-output theory system, with input and output fields $a_{\text{in}}$ and $a_{\text{out}}$, intracavity field operator $a$, and bath input and output fields $a_\text{in}$ and $a_\text{out}$. The input field is coupled to the cavity at rate $\kappa$, and the cavity to the bath at rate $\gamma$. The accompanying circuit is coloured according to the correspondence with the associated fields and coupling constants. The bath continuum is coupled to the circuit via the resistor.
	}
\end{figure}

The field operators $a$, $\Phi$ and $Q$ and Hamiltonian $H_\kir$ so far describe the intracavity field dynamics. In the experimental setting we stimulate the resonator with an input field operator $a_{\text{in}}$ and measure a reflected response $a_{\text{out}}$ that enter and exit the cavity via the coupling circuit. For example, a Vector Network Analyzer (VNA) measures the reflection parameter $S_{11}(\omega) = \langle a_{\text{in}} \rangle / \langle a_{\text{out}} \rangle$. To obtain a classical description of the microwave response we would typically adopt a scattering matrix approach \cite{pozar_microwave_2012}. The below derivation follows that presented in Ref.~\citenum{boutin_2015}, and is reproduced here for completeness.

Input-output theory, developed by Gardiner and Collett \cite{gardiner_1985}, extends the scattering matrix formalism to the quantum regime. Let $H$ be the Hamiltonian written in terms of the creation and annihilation operators $a^\dagger$ and $a$, where $H$ is coupled to the bath at rate $\gamma$, used to model the losses in the system, and input field $a_{\text{in}}$ at rate $\kappa$ (see~\cref{fig:io_model}). We write down the following Heisenberg picture master equation to describe the system:
\begin{equation}
	\frac{\partial a(t)}{\partial t} = \frac{[a, H]}{i\hbar} - \bar{\kappa}a(t) + \sqrt{\kappa}a_{\text{in}} + \sqrt{\gamma}b_\text{in}(t)
	\label{eqn:io_master}
\end{equation}
where $\bar{\kappa} = (\gamma + \kappa)/2$. The output field operator $a_{\text{out}}$ is then given by the input-output relation:
\begin{equation}
	a_{\text{out}}(t) - a_{\text{in}}(t) = \sqrt{\kappa}a(t)
	\label{eqn:io_rel}
\end{equation}

Consider now the linear $\lambda/4$ resonator Hamiltonian provided in~\cref{eq:Hlin}, truncated to the fundamental mode. We re-write~\cref{eqn:io_master} in the Fourier domain using:
\begin{equation}
    a[\omega] = \frac{1}{\sqrt{2\pi}} \int_{-\infty}^\infty e^{i\omega t} a(t) d\omega  
\end{equation}
which gives: 
\begin{align}
	-i\omega a[\omega] &= -\frac{i}{\hbar}\big[a[\omega], \hbar\omega_0 a^\dagger[\omega] a[\omega]\big] - \bar{\kappa}a[\omega] + \sqrt{\kappa}a_{\text{in}}[\omega] + \sqrt{\gamma}b_\text{in}[\omega]\\
	-i\omega a &= -i\omega_0a - \bar{\kappa}a + \sqrt{\kappa}a_{\text{in}} + \sqrt{\gamma}b_\text{in}
\end{align}

Substituting for $a$ using the input-output relation (\cref{eqn:io_rel}) yields the output field operator in terms of the input and bath fields:
\begin{align}
	-i\omega(a_{\text{out}} - a_{\text{in}}) &= -(\bar{\kappa} + i\omega_0)(a_\text{out} - a_\text{in}) + \kappa a_{\text{in}} + \sqrt{\frac{\gamma}{\kappa}}b_\text{in} \\
	\Rightarrow a_\text{out} &= \bigg(\frac{\kappa}{\bar{\kappa} - i(\omega - \omega_0)} - 1\bigg)a_\text{in} + \frac{\sqrt{\gamma/\kappa}}{\bar{\kappa} - i(\omega - \omega_0)} b_\text{in}
\end{align}

Treating the bath input field $b_\text{in}$ as a thermal state such that $\langle b_\text{in} \rangle = 0$, we retrieve the expression for the reflection parameter:
\begin{align}
	S_{11}[\omega]
		&= \frac{\langle a_\text{out}[\omega]\rangle}{\langle a_\text{in}[\omega]\rangle}\\
		&= \frac{\kappa}{\bar{\kappa} - i(\omega - \omega_0)} - 1 \label{eqn:s11_reflection}
\end{align}

We can apply the same mathematics to the idealized KIPA Hamiltonian (\cref{eq:kipadpa}). We first write the master equation in the Fourier domain, as before~\cite{boutin_2017}:
\begin{align}
	-i\omega a 
		&= -\bigg(\bar{\kappa}a + i\Delta a + i\frac{\xi}{2}\big[a, a^{\dagger2}\big] + i\frac{\xi^*}{2}\big[a, a^{2}\big]\bigg) + \sqrt{\kappa} a_\text{in} + \sqrt{\gamma}b_\text{in}\\
	-i\omega a 
		&= -((\bar{\kappa} + i\Delta)a + i\xi a^\dagger) + \sqrt{\kappa} a_\text{in} + \sqrt{\gamma}b_\text{in} \label{eqn:dpa_io_partial}
\end{align}

Next, we take the Hermitian conjugate of both sides. Note that in the Fourier domain $(a[\omega])^\dagger = a^\dagger[-\omega]$. To simplify notation the frequency reversal is implied. We find:
\begin{equation}
	i\omega a^\dagger = -((\bar{\kappa} - i\Delta)a^\dagger - i\xi^*a) + \sqrt{\kappa}a_\text{in}^\dagger + \sqrt{\gamma}b_\text{in}^\dagger
\end{equation}

Combined with~\cref{eqn:dpa_io_partial}, we obtain the matrix equation:
\begin{align}
	i\omega \begin{pmatrix} -a \\ a^\dagger \end{pmatrix}
		&= \begin{pmatrix}
			-i\Delta - \bar{\kappa} & - i\xi \\
			i\xi^* & i\Delta - \bar{\kappa}
		\end{pmatrix}
		\begin{pmatrix} a \\ a^\dagger \end{pmatrix}
		+ \sqrt{\kappa}
		\begin{pmatrix} a_\text{in} \\ a_\text{in}^\dagger \end{pmatrix}
		+ \sqrt{\gamma}
		\begin{pmatrix} b_\text{in} \\ b_\text{in}^\dagger \end{pmatrix}\\
	\Rightarrow \begin{pmatrix} a \\ a^\dagger \end{pmatrix}
		&= -\sqrt{\kappa}\begin{pmatrix}
			-i\Delta + \bar{\kappa} + i\omega & -i\xi \\
			i\xi^* & i\Delta - \bar{\kappa} - i\omega
		\end{pmatrix}^{-1}
		\Bigg[
		\begin{pmatrix} a_\text{in} \\ a_\text{in}^\dagger \end{pmatrix}
		+ \sqrt{\frac{\gamma}{\kappa}}
		\begin{pmatrix} b_\text{in} \\ b_\text{in}^\dagger \end{pmatrix}
		\Bigg] \\
		&= -\frac{\sqrt{\kappa}}{D[\omega]} 
		\begin{pmatrix}
			-i\Delta + \bar{\kappa} + i\omega & -i\xi \\
			i\xi^* & i\Delta - \bar{\kappa} - i\omega
		\end{pmatrix}
		\Bigg[
		\begin{pmatrix} a_\text{in} \\ a_\text{in}^\dagger \end{pmatrix}
		+ \sqrt{\frac{\gamma}{\kappa}}
		\begin{pmatrix} b_\text{in} \\ b_\text{in}^\dagger \end{pmatrix}
		\Bigg]
\end{align}
where $D[\omega] = \Delta^2 + (\bar{\kappa} - i\omega)^2 - |\xi|^2$. Substituting the input-output relation (\cref{eqn:io_rel}) gives the input-output equation for the ideal DPA \cite{boutin_2017}:
\begin{equation}
	a_\text{out}[\omega] = g_S[\omega] a_\text{in}[\omega] + g_I[\omega]a_\text{in}^\dagger[-\omega] + \sqrt{\frac{\gamma}{\kappa}}\bigg[(g_S[\omega] + 1) b_\text{in}[\omega] + g_I[\omega]b_\text{in}^\dagger[-\omega]\bigg]
	\label{eqn:boutin_io}
\end{equation}
where we make the frequency reversal explicit, and define the signal and idler gains:
\begin{equation}
	g_S[\omega] = \frac{\kappa\bar{\kappa} - i\kappa(\Delta + \omega)}{D[\omega]} - 1, \quad \quad g_I[\omega] = \frac{-i\xi\kappa}{D[\omega]}
	\label{eqn:boutin_refltion_params}
\end{equation}

\section{Phase Insensitive Amplification} \label{sec:phase_insens_gain}
\begin{figure}
    \centering
    \includegraphics[width=\textwidth]{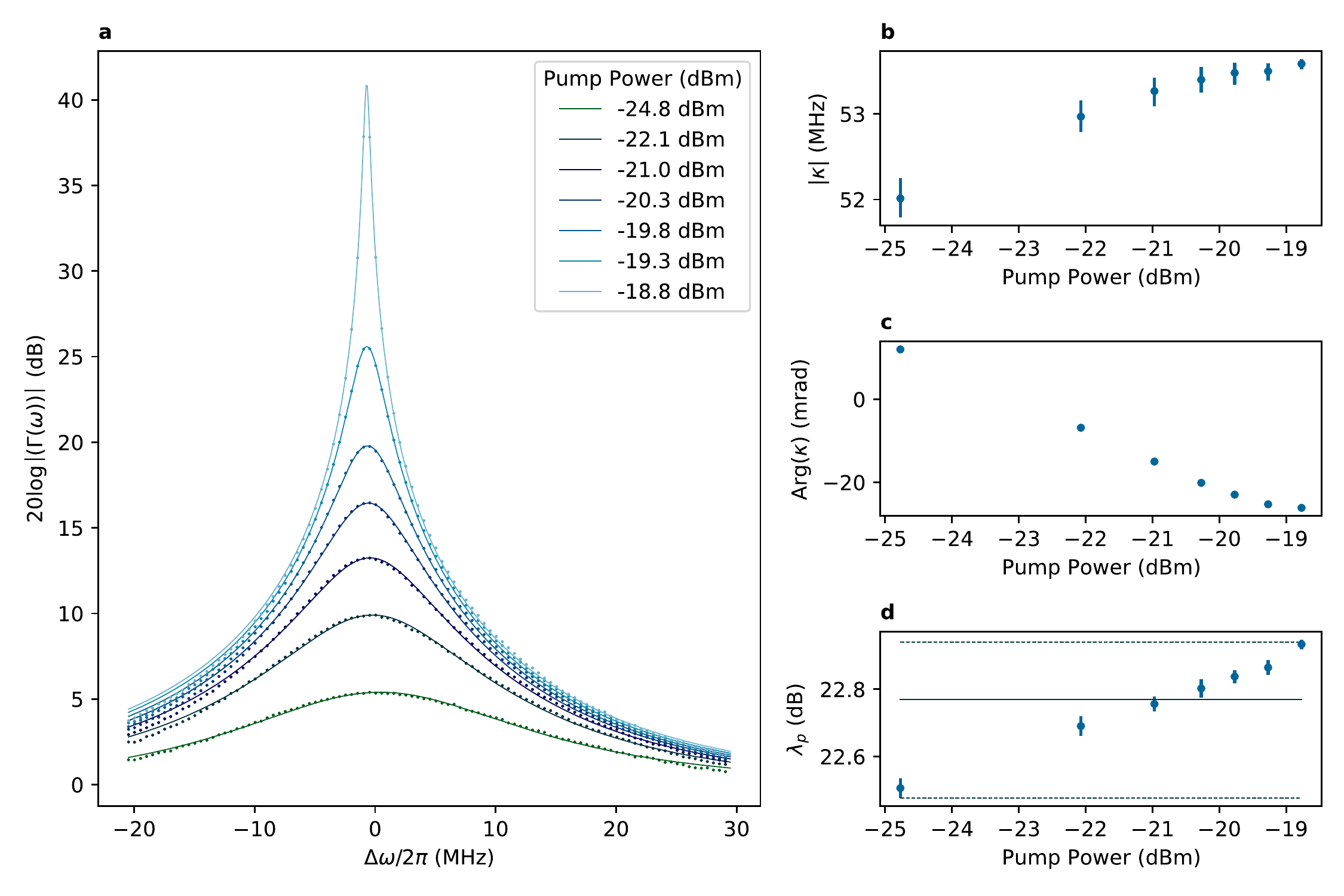}
    \caption{\label{fig:phase_insenitive_gain}
        (a) Phase insensitive gain as a function of frequency $\omega = \Delta \omega + \omega_p/2$ for different pump powers (circles). Traces are labelled by the pump power at the cavity input. The fitted theoretical frequency response is plotted (solid lines). The parameters $|\kappa|$, $\text{Arg}(\kappa)$ and the pump line transmittance $\lambda_p$ were the only free parameters. (b) $|\kappa|$ as a function of the estimated pump current in the device (circles), with linear fit (solid line). (c) $\text{Arg}(\kappa)$ as a function of pump power. The dashed line is a guide for the eye.
    }
\end{figure}

Phase insensitive gain is readily measured with a vector network analyzer (VNA) once an appropriate pump tone and bias current are applied to the device concurrently. For the measurements in Fig.~2a of the main text, we chose a pump frequency of $\omega_p/2\pi = 14.381\,\text{GHz} \approx \omega_0/\pi$, close to twice the resonant frequency of $\omega_0 = 7.1924$ GHz for the bias current $I_\dc = 0.834$ mA. Ideally, the KIPA should be operated at precisely $\Delta = 0$, or $\omega_p \approx 2(\omega_0 + \delta_\dc + \delta_p)$ for maximal gain. However, in our experiments we optimise the pump frequency for gain at a fixed pump power and bias current, arriving at a close to optimal pump configuration. A fixed pump frequency of $\omega_p/2\pi = 14.381$ GHz is used throughout the experiments, despite the expected shift in resonance (\cref{eq:Hamparam}).

The VNA supplies a signal tone, which is swept about $\omega_p/2$, while the reflected response from the KIPA is measured. Because the magnitude response of the KIPA is approximately flat, we measure gain by taking the difference between the response with the pump on and the pump off, depicted in Fig.~2 at different pump powers. The KIPA produces an amplified signal tone $\omega_s$ at its output, along with an idler at $\omega_i$ such that energy is conserved in the 3WM process $\omega_p = \omega_s + \omega_i$. Phase insensitive gain occurs when $\omega_s = \omega_p/2 + \Delta\omega$ with $|\Delta\omega|$ exceeding the bandwidth resolution of the measurement. Gain increases with the pump power and is found to be in excess of $40\,\text{dB}$ before the KIPA crosses the threshold where spontaneous parametric oscillations occur (see~\cref{sec:PO}).

\subsection{Gain Feature Fits}\label{sec:gain_features}
To characterise the line-shape of the non-degenerate amplification features in the main text, we define the reflection parameter $\Gamma(\omega)$, which is simply the signal gain (\cref{eqn:boutin_refltion_params}) written in the laboratory frame (i.e. $g_S[\omega] \to \Gamma(\omega)$ with $\omega \to \omega - \omega_p/2$) \cite{boutin_2017}:
    \begin{equation}
        \Gamma(\omega) = 
            \frac{\kappa(\kappa + \gamma)/2 + i\kappa(\Delta + \omega - \omega_p/2)} {\Delta^2 + \big[(\kappa + \gamma)/2 + i(\omega - \omega_p/2)\big]^2 - |\xi|^2} - 1
        \label{eqn:S11}
    \end{equation}

To fit the data in Fig.~2, we adopt a complex coupling rate in the reflection model $\Gamma(\omega)$: $\kappa \in \mathbb{R} \to \kappa \in \mathbb{C}$, with complex phase $\text{Arg}(\kappa)$. A complex quality factor may be used to model an asymmetric response that occurs due to an impedance mismatch across the coupling circuit where reflections at the coupler interfere with photons exiting the resonator \cite{khalil_2012,probst_2015}.

The pump current in our device is not precisely known. We simulate an impedance of $Z_0 = 118~\Omega$ for the $\lambda/4$ resonator (using the software Sonnet) and introduce a loss parameter $\lambda_p$ that quantifies the amount of pump power transmitted from room temperature down to the sample such that $I_p^2 = 2\lambda_p P_\text{pump} / Z_0$, where $P_\text{pump}$ is the pump power at microwave source. 

We may predict the parameter $\Delta = \omega_0 + \delta_\dc + \delta_p - \omega_p/2$ from our theory (\cref{eq:Hamparam}a,b) as $\omega_p$ and $I_\dc$ are known, and we have measured $\omega_0$ and $I_*$. Further, \cref{eq:Hamparam}d allows us to also predict $|\xi|$ as a function of the pump current in the sample. To further constrain the model we assume the quality factor $Q_i = \omega_0/\gamma$ to be $10^5$ for all pump powers. We base this estimate on the internal quality factors observed in similar devices (e.g. the device measured in Fig.~S\ref{fig:BA13_fits}), and note that in the over-coupled regime in which the KIPA operates $Q_i$ has minimal impact on the predicted gain as $\kappa + \gamma \approx \kappa$. We are left with a model where the only free parameters are $|\kappa|$, $\text{Arg}(\kappa)$ and $\lambda_p$. The fit results are shown in Fig.~S\ref{fig:phase_insenitive_gain}.

We observe that $\kappa$ varies from $\sim 52$ MHz to $\sim 54$ MHz, corresponding to an average coupling quality factor of $Q_c \approx 135$. The RMS pump current increases the kinetic inductance and thus modifies the coupling circuit (i.e. it changes the impedance step in the band stop filter), which might explain the pump power dependent coupling rate $\kappa$. A weak drift ($\sim 30$ mrad) in the phase of the coupling rate was necessary to fit the data (see Fig.~S\ref{fig:phase_insenitive_gain}c). This is not unreasonable, as a small shift in the cavity and coupling circuit impedances due to the pump current will influence any reflections that occur at the cavity input.

From the fits, we extract an average pump attenuation of $-10\text{log}(\lambda_p) = 22.8$~dB. At room temperature the measured loss of the lines and components is $\sim 30$ dB. Since the line and component loss is expected to reduce at cryogenic temperatures, this fitted average pump attenuation is realistic. We also note that the extracted pump attenuation increases marginally as the pump power increases, raising by $\sim 0.4$~dB over the range of powers explored (see Fig.~S\ref{fig:phase_insenitive_gain}d). This could be an indication that the pump becomes slightly depleted as the gain raises \cite{sivak_2019}. 

Overall, we find excellent quantitative agreement with our theory, and are able to predict the observed gain curves from the KIPA Hamiltonian (Eqs.~(\ref{eq:HKIPArot}-\ref{eq:Hamparam})) derived in \cref{sec:Ham}.

\subsection{Gain Bandwidth Product}
\begin{figure}
    \centering
    \includegraphics[width=0.7\textwidth]{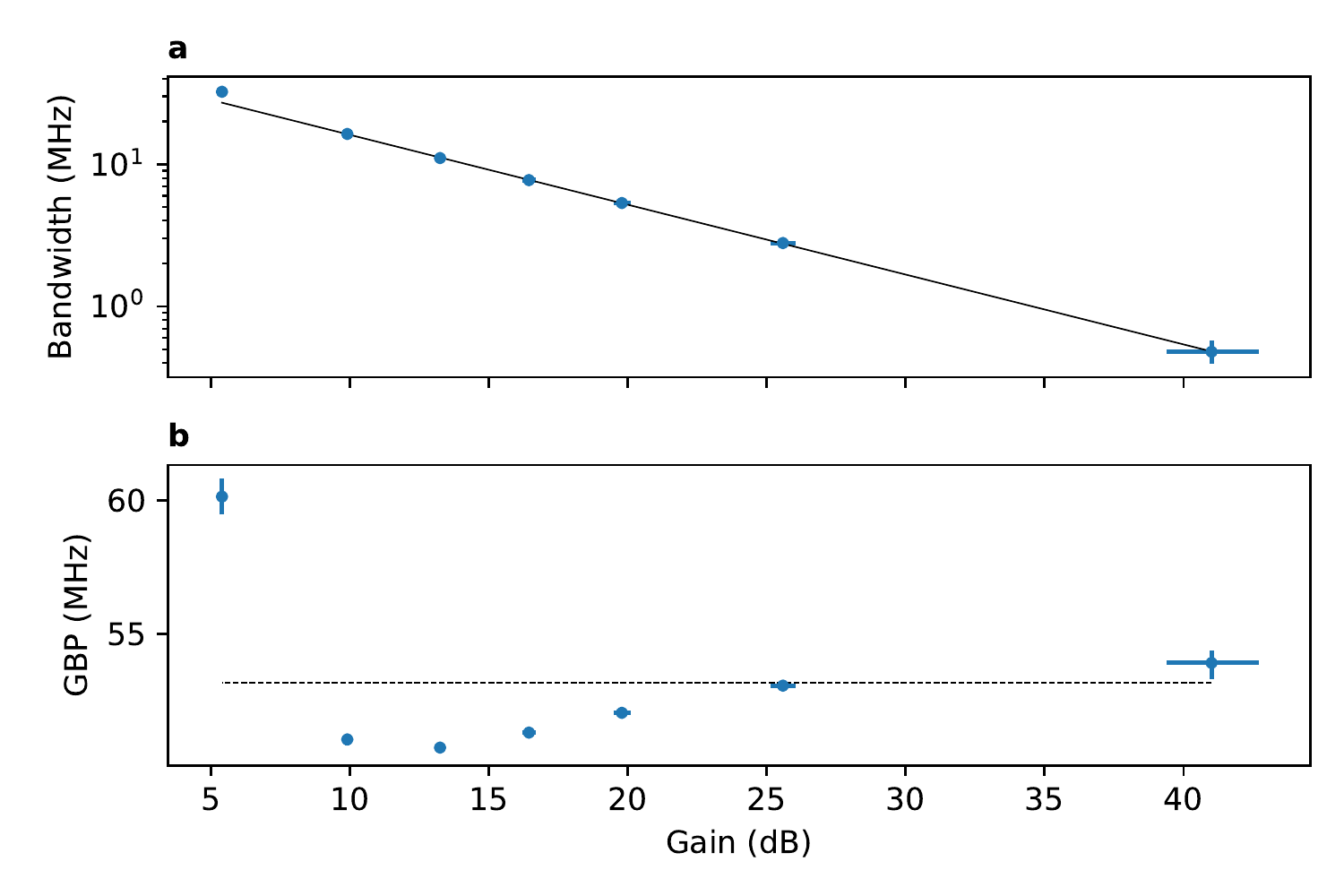}
    \caption{\label{fig:gbp}
        (a) Bandwidth vs peak phase insensitive gain. The solid black line is a log-linear fit to the bandwidth as a function of gain in dB. (b) Gain Bandwidth Product (GBP) vs peak phase insensitive gain (bottom). The dashed black line is the average GBP across all gains.
    }
\end{figure}

From the fits to the amplification features depicted in Fig.~S\ref{fig:phase_insenitive_gain}, we can extract the Gain Bandwidth Product (GBP), defined by the product of the peak phase insensitive amplitude gain $G$ and the bandwidth when the amplitude gain drops to $G/\sqrt{2}$ \cite{zhou_2014}. We find that the GBP of the KIPA shows good consistency across the different pump powers, as evidenced by the highly linear log-log plot of the gain and bandwidth (see Fig.~S\ref{fig:gbp}), and we extract an average GBP of $53(7)$ MHz.

\subsection{Parametric Self Oscillations}\label{sec:PO}
Increasing the pump current $I_p$, and hence $\xi$, will not increase the gain indefinitely. Past a certain threshold, the device enters the regime of parametric self-oscillation and ceases to behave as an amplifier \cite{dykman_2012}. Pumped at twice the resonant frequency, the cavity spontaneously produces photons at $\omega_0$ that grow rapidly in number. Competition from system nonlinearities eventually limit growth, resulting in a fixed power $\omega_0$ tone at steady state. 

Although, we do not study the KIPA in the self-oscillation regime in this work, we use our theory to predict the range of pump currents at which the KIPA behaves as a parametric amplifier. The parametric oscillation threshold corresponds to the zero crossing of the denominator of $|\Gamma(\omega)|$. At the point of maximum phase sensitive amplification, spontaneous oscillations occur when $|\xi|^2 \ge \Delta^2 + (\kappa + \gamma)^2/4$. Using our theory along with the coupling rate $|\kappa|$ extracted from the fits depicted in Fig.~S\ref{fig:phase_insenitive_gain}, we can predict the pump current at which parametric self-oscillation occurs. We assume a real coupling rate $\kappa$ to simplify the analysis.

\begin{figure}
    \centering
    \includegraphics[width=0.7\textwidth]{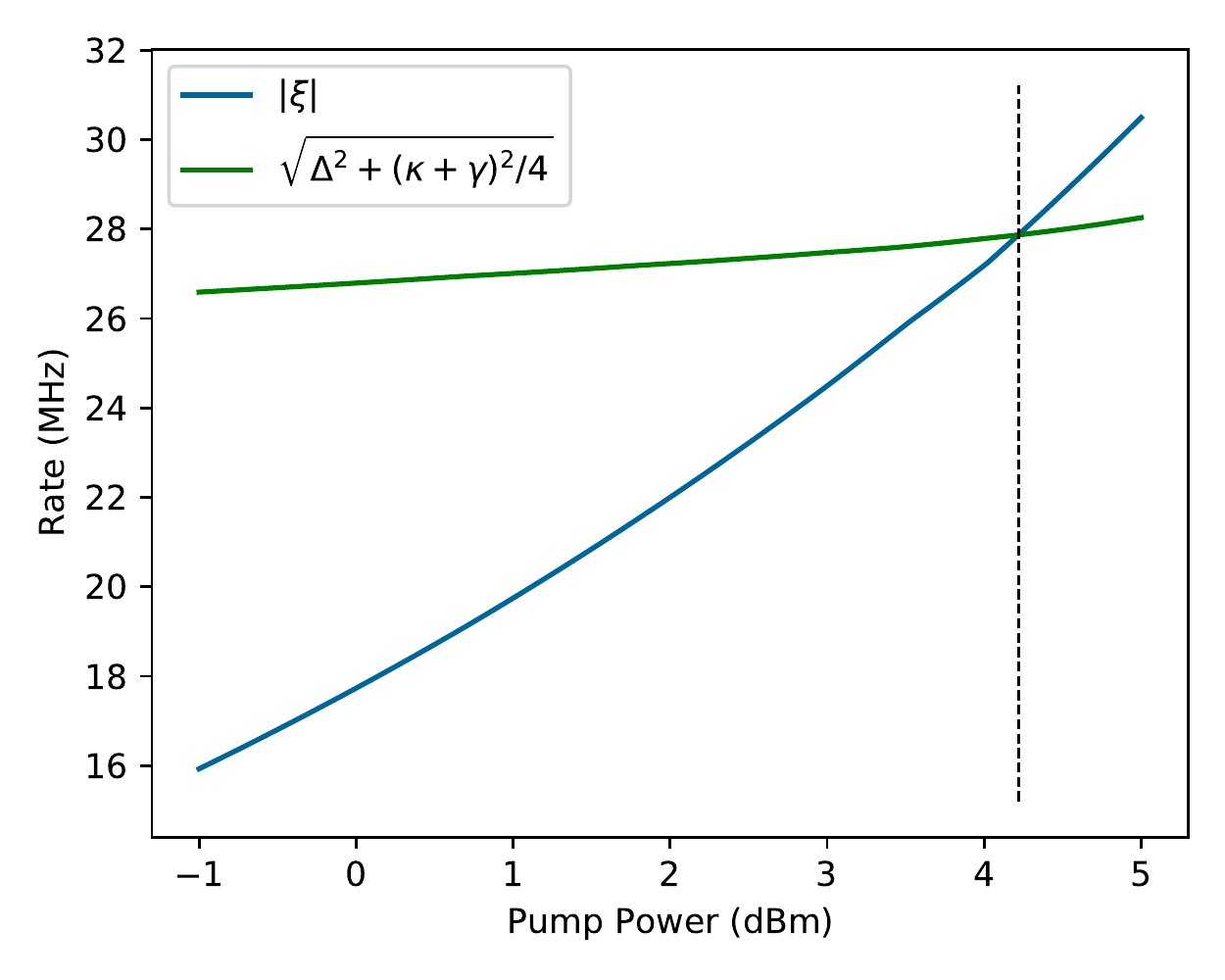}
    \caption{\label{fig:self_osc_threshold}
        The rate $|\xi|$ and parametric self oscillation threshold $\sqrt{\Delta^2 + (\kappa + \gamma)^2/4}$ versus the pump power at the microwave source output. The parametric self oscillation threshold occurs at the intersection of these curves indicated by the black dashed line at $P_\text{pump} = 4.22$ dBm.
    }
\end{figure}

Fig.~S\ref{fig:self_osc_threshold} depicts the predicted $|\xi|$ as a function of pump power alongside the predicted threshold of parametric self-oscillation: $\sqrt{\Delta^2 + (\kappa + \gamma)^2/4}$. The threshold increases with the pump power due to the pump dependent detuning $\delta_p$, which increases $\Delta^2$ as the pump current becomes larger. The curves intersect at a pump power of $P_\text{pump} = 4.22$ dBm referred to the output of our microwave source.

We found in practice that the KIPA would self-oscillate beyond a pump power of 4.10~dBm, demonstrating an excellent quantitative agreement between experiment and theory.

\section{Phase Sensitive Amplification} \label{sec:phase_sens_gain}
\begin{figure}
    \centering
    \includegraphics[width=\textwidth]{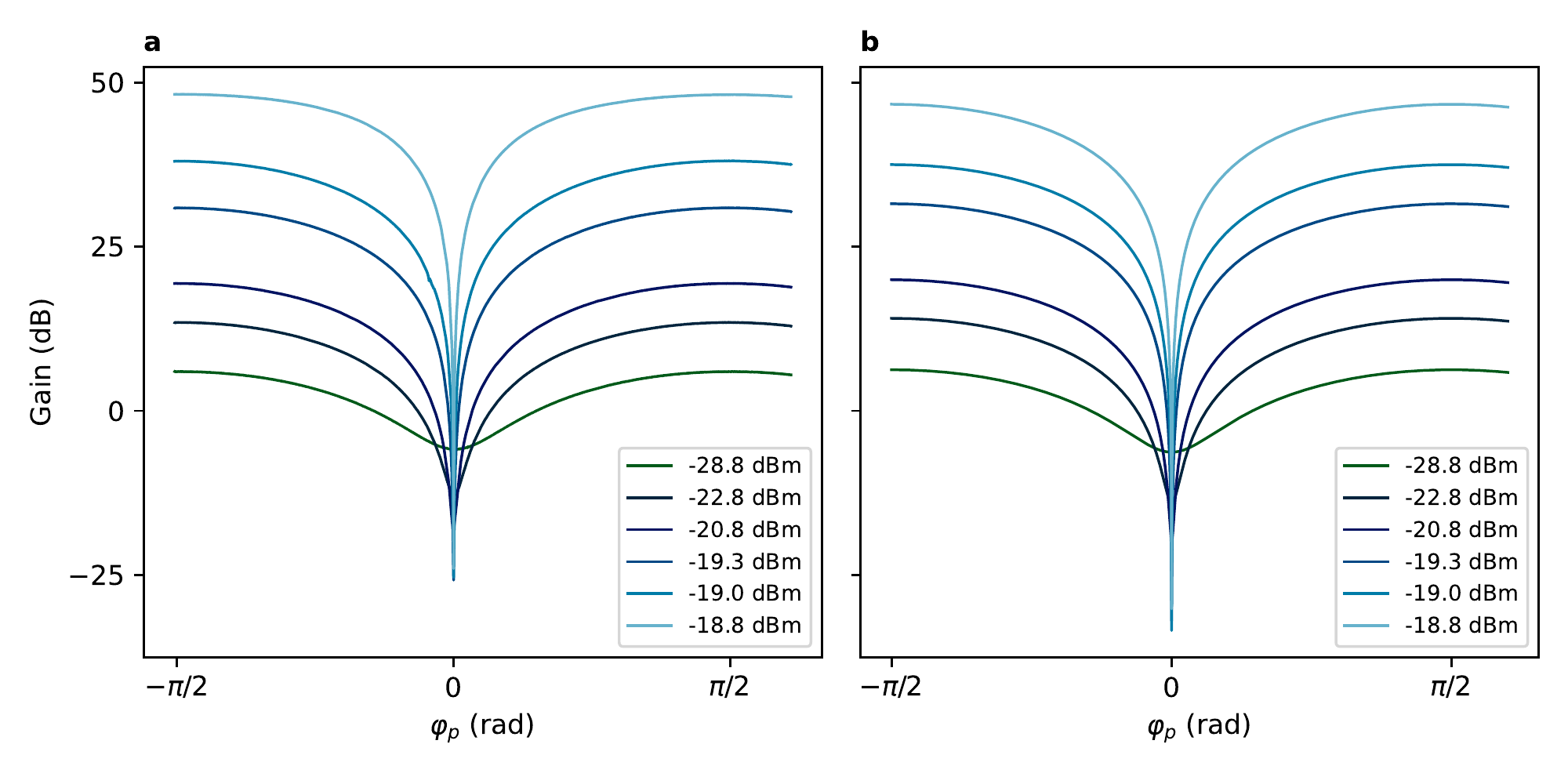}
    \caption{\label{fig:maccas}
        Measured gain (a) and theoretical gain (b) as a function of the pump/signal phase for an input signal of frequency $\omega_s = \omega_p/2$. Traces are labelled by the pump power at the cavity input (see legend). a slight discrepancy between theory and experiment at the highest pump power (-18.8 dBm) is due to an increasing sensitivity of the peak theoretical gain to coupling rate $|\kappa|$.
    }
\end{figure}
When applying a signal tone at half the pump frequency $\omega_s = \omega_p/2$, the KIPA enters the degenerate mode of operation, producing phase sensitive gain as the signal and idler tones interfere. From the input-output equations (see \cref{eqn:boutin_refltion_params}), the rotating frame gain is: $\Gamma(\varphi_p) = g_S[0] + g_I[0]$. More explicitly, the phase sensitive gain is:
\begin{align}
    |\Gamma(\varphi_p)| 
        &= \bigg|\frac{\kappa(\kappa + \gamma)/2 + i\kappa\Delta + i\kappa|\xi|e^{-j\varphi_p}}{\Delta^2 + (\kappa + \gamma)^2/4 - |\xi^2|} - 1\bigg| \\
        &= \sqrt{\bigg[\frac{\kappa((\kappa + \gamma)/2 - |\xi|\sin(\varphi_p))}{\Delta^2 + (\kappa + \gamma)^2/4 - |\xi|^2} - 1\bigg]^2 + \bigg[\frac{\kappa(\Delta + |\xi|\cos(\varphi_p))}{\Delta^2 + (\kappa + \gamma)^2/4 - |\xi|^2}\bigg]^2}
        \label{eqn:phase_sens_gain}
\end{align}
where we separate out the modulus of $\xi$ and its argument corresponding to the pump phase $\varphi_p$ (see \cref{eq:Hamparam}d). From \cref{eqn:phase_sens_gain} we observe that the KIPA gain is sensitive to the pump phase $\varphi_p$.

Experimentally, we observe phase sensitive amplification by modulating the phase of a signal tone which has a frequency of $\omega_p / 4\pi = 7.1905$ GHz. As $\varphi_p$ represents the phase difference between the signal and the pump, phase modulation of either tone will allow us to characterise the phase sensitive gain. Fig.~(2)b of the main text (reproduced here in Fig.~S\ref{fig:maccas}a) depicts the gain of the KIPA as a function of pump phase, where up to $26\,\text{dB}$ of deamplification and close to $50\,\text{dB}$ of amplification are observed. Compared to phase insensitive amplification, additional gain is observed in degenerate mode due to the constructive interference that occurs between the signal and idler. The traces are aligned such that the point of maximum deamplification occurs for $\varphi_p = 0$.

Fig.~S\ref{fig:maccas}b shows the phase sensitive gain predicted by our theory (\cref{eqn:phase_sens_gain}), where we use interpolated data from the fitted $\kappa$ points in Fig.~S\ref{fig:phase_insenitive_gain}, the extracted pump loss $\lambda_p = 22.8$ dB, and the pump current dependent expressions for $\xi$ and $\delta \omega$ from our Hamiltonian derivation (\cref{sec:Ham}). We find excellent agreement with theory for the amplification regions of each pump power. On the other hand, the theory predicts greater deamplification than is observed experimentally for the three highest pump powers. To obtain the data plotted in Fig.~S\ref{fig:maccas}a, significant averaging was required to reduce the noise. We believe that the maximum deamplification of $26\,\text{dB}$ measured is partially limited by our ability to resolve the sharp gain feature at $\varphi_{p} = 0$, which is highly sensitive to instrumental phase noise and slow phase drifts between the signal and pump. Reflections may also limit the observed deamplification, as discussed in \cref{sec:sqz_tform_theory}. 

\subsection{1dB Compression Point}
The 1 dB compression point of the KIPA is characterised in phase sensitive mode. After calibrating the phase of the pump to achieve maximum amplification (i.e. $\varphi_{p} \approx \pi/2$), we characterise the degenerate $1$ dB compression point of the KIPA by increasing the signal power until the gain drops by $1$ dB, as is presented in Fig.~2c of the main text. For $\sim 20\,\text{dB}$ of phase sensitive gain, we find a compression power of $-49.5(8)\,\text{dBm}$ at the KIPA output, comparable to the compression performance of kinetic inductance travelling wave amplifiers \cite{eom_2012,vissers_2016,malnou_2021}. Our HEMT is expected to saturate for approximate input powers of $\sim -46$ dBm \cite{LNF_2021}. Factoring in the loss between the KIPA and the HEMT, we are unable to rule out the possibility that the measured the 1 dB compression point is limited by the HEMT, and that the dynamic range of the KIPA is indeed higher.

\section{Squeezing Transformation of the DPA} \label{sec:sqz_tform_theory}
Re-writing the input-ouptput relation (\cref{eqn:boutin_io}) in the degenerate case ($\omega=0$), we find \cite{boutin_2017}:
\begin{align}
    a_\text{out} &=
        g_s a_\text{in} + g_i a^\dagger_\text{in} +
        \sqrt{\frac{\gamma}{\kappa}}\big((g_s + 1)b_\text{in} + g_i b_\text{in}^\dagger\big)\\
    a^\dagger_\text{out} &=
        g_s^* a^\dagger_\text{in} + g_i^* a_\text{in} +
        \sqrt{\frac{\gamma}{\kappa}}\big((g_s^* + 1)b^\dagger_\text{in} + g_i^* b_\text{in}\big)
\end{align}
giving the output quadrature relations:
\begin{align}
    I_\text{out} &= \frac{1}{2}(a_\text{out}^\dagger + a_\text{out})
            = \frac{1}{2}\bigg[
                \epsilon a_\text{in} + \epsilon^* a^\dagger_\text{in} +
                \sqrt{\frac{\gamma}{\kappa}}\big((\epsilon + 1)b_\text{in} + (\epsilon^* + 1)b_\text{in}^\dagger\big)
            \bigg]\label{eqn:Iout}\\
    Q_\text{out} &= \frac{i}{2}(a_\text{out}^\dagger - a_\text{out})
            = \frac{i}{2}\bigg[
                \epsilon'^* a^\dagger_\text{in} - \epsilon' a_\text{in} +
                \sqrt{\frac{\gamma}{\kappa}}\big((\epsilon'^* + 1)b^\dagger_\text{in} - (\epsilon' + 1) b_\text{in}\big)
            \bigg]\label{eqn:Qout}
\end{align}
where $\epsilon = g_s + g_i^*$ and $\epsilon' = g_s - g_i^*$. Using the
identities:
\begin{align}
    \frac{1}{2}\big[\beta^* a^\dagger + \beta a \big]
        &= \text{Re}(\beta)I - \text{Im}(\beta)Q\\
    \frac{i}{2}\big[\beta^* a^\dagger - \beta a \big]
        &= \text{Re}(\beta)Q + \text{Im}(\beta)I
\end{align}
where $\beta$ is an arbitrary complex number (such as $\epsilon$ or $\epsilon'$), we arrive at a set of linear equations for the output field quadratures:
\begin{equation}
    \begin{pmatrix}
        I_\text{out} \\
        Q_\text{out}
    \end{pmatrix}
    = A_G
    \begin{pmatrix}
        I_\text{in} \\
        Q_\text{in}
    \end{pmatrix}
    + \sqrt{\frac{\gamma}{\kappa}} \big( A_G + 1\big)
    \begin{pmatrix}
        I_{b} \\
        Q_{b}
    \end{pmatrix} \label{eqn:io_matrix}
\end{equation}
where $I_b$ and $Q_b$ are the quadratures of the bath field. As a function of the pump phase $\varphi_p$, the affine transformation of the quadratures $A_G$ is given by:
\begin{align}
    A_G(\varphi_p) &=
    \begin{pmatrix}
        \text{Re}(\epsilon) & -\text{Im}(\epsilon) \\
        \text{Im}(\epsilon') & \text{Re}(\epsilon')
    \end{pmatrix}\\
    &= \frac{\kappa}{\Delta^2 + (\kappa + \gamma)^2/4 - |\xi|^2}
    \begin{pmatrix}
        (\kappa + \gamma)/2 - |\xi|\sin(\varphi_p) & -|\xi|\cos(\varphi_p) + \Delta\\
        -|\xi|\cos(\varphi_p) - \Delta & (\kappa + \gamma)/2 + |\xi|\sin(\varphi_p)
    \end{pmatrix} - 1 \label{eqn:matrix}
\end{align}

The pump phase $\varphi_p$ has the effect of rotating the basis of the transformation. In fact, one can show that $A_G(\varphi_p) = R^T(\varphi_p)A_G(0)R(\varphi_p)$ where $R(\theta)$ is the standard $2\times2$ rotation matrix.

As the bath is a thermal state, taking the expectation of both sides of \cref{eqn:io_matrix} gives the simple expression:
\begin{equation}
    \begin{pmatrix}
        \langle I_\text{out} \rangle \\
        \langle Q_\text{out} \rangle
    \end{pmatrix}
    = A_G(\varphi_p)
    \begin{pmatrix}
        \langle I_\text{in} \rangle \\
        \langle Q_\text{in} \rangle
    \end{pmatrix}
\end{equation}

\begin{figure}
    \centering
    \includegraphics[width=0.6\textwidth]{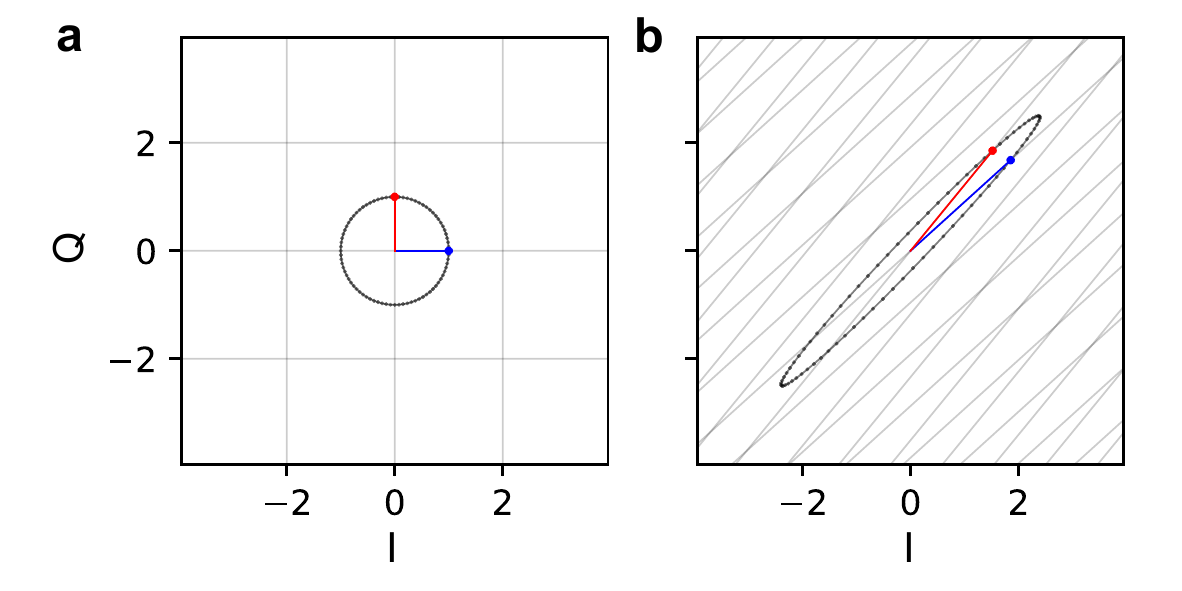}
    \caption{\label{fig:affine_0}
        Visualisation of the linear transformation $A_G(0)$ acting on points on the unit circle. The red (blue) points/lines correspond to the standard basis vectors in $V$. (a) Unit circle before the transformation $A_G(0)$. (b) Unit circle after the transformation.
    }
\end{figure}

Assuming $\Delta = 0$ we find $A_G \to 1$ in the limit that $|\xi| \to 0$, as expected. Conversely, if $\Delta \neq 0$ then $A_G$ is an affine transformation that will always mix the input quadratures to some degree, limiting the achievable squeezing for a given $\xi$. Fig.~S\ref{fig:affine_0} illustrates the mapping of points on the unit circle $(I, Q)^T = (\sin(\phi), \cos(\phi))^T$ in the vector space $V \in \mathbb{R}^2$ by the linear transformation $A_G(0): V \to W$. Setting $\varphi_p = 0$ yields a mapping where the standard unit vectors in $V$ do not in general map to the standard unit vectors in $W$, nor do they correspond to the principal axes of the elliptical output state.

\begin{figure}
    \centering
    \includegraphics[width=0.6\textwidth]{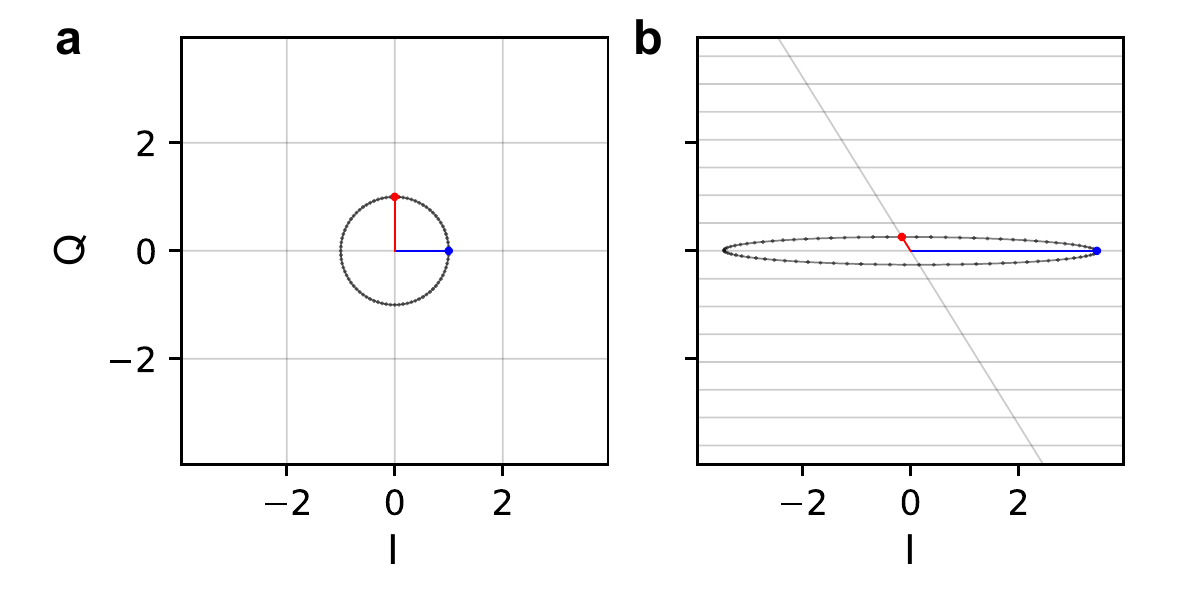}
    \caption{\label{fig:affine_horz}
        Visualisation of the linear transformation $A_G(\pi/2 - \arccos(-\Delta/|\xi|))$ acting on points on the unit circle. The red (blue) points/lines correspond to the standard basis vectors in $V$. (a) Unit circle before the transformation $A_G(\pi/2 - \arccos(-\Delta/|\xi|))$. (b) Unit circle after the transformation.
    }
\end{figure}

We may align the axis of amplification along $Q$, as depicted in Fig.~S\ref{fig:affine_horz}, by choosing $\varphi_p = \pi/2 - \arccos(-\Delta/|\xi|)$. Note that in Fig.~S\ref{fig:affine_horz}, we deliberately set $\Delta \neq 0$ to illustrate the fact that orthogonal vectors in $V$ do not necessarily map to orthogonal vectors in $W$. On the other hand, when $\Delta=0$, the optimal angle of rotation will correspond to $\varphi_p = 3\pi/2$ giving a strictly diagonal matrix $A_{G}(\varphi_p)$ with partial diagonal elements $(\kappa + \gamma)/2 + |\xi|$ and $(\kappa + \gamma)/2 - |\xi|$, such that orthogonality is preserved. Degenerate amplification increases as $|\xi|$ approaches the asymptote of self oscillation ($|\xi|^2 = \Delta^2  + (\kappa + \gamma)^2/4$), while simultaneously, deamplification approaches 0.

The expression for gain as a function of the pump phase $\varphi_p$ is given by:
\begin{align}
    g(\varphi_p)
        &= \frac{|| (\langle I_\text{out} \rangle, \langle Q_\text{out} \rangle)^T ||}{|| (\langle I_\text{in} \rangle, \langle Q_\text{in} \rangle)^T ||} \\
        &= \sqrt{\frac{(I_\text{in}g_{11}(\varphi_p) + Q_\text{in}g_{12}(\varphi_p))^2 + (I_\text{in}g_{21}(\varphi_p) + Q_\text{in}g_{22}(\varphi_p))^2}{I_\text{in}^2 + Q_\text{in}^2}}
\end{align}
where $g_{ij}$ are the matrix elements of $A_G(\varphi_p)$. This corresponds exactly with the expression for phase sensitive gain provided earlier in \cref{eqn:phase_sens_gain}.

\subsection{Reflections with the Predicted DPA Transformation}
The ellipses depicted in Fig.~3 of the main text are not simply a result of the squeezing transformation applied to coherent inputs of fixed magnitude. Because our setup is not perfectly impedance matched, reflections will occur at the input to the KIPA (e.g. from the PCB and input connector) that superimpose on the squeezing transformation. Although these reflections only account for a small percentage of the detected signal, they become considerable as the deamplification increases.

\begin{figure}
    \centering
    \includegraphics[width=\textwidth]{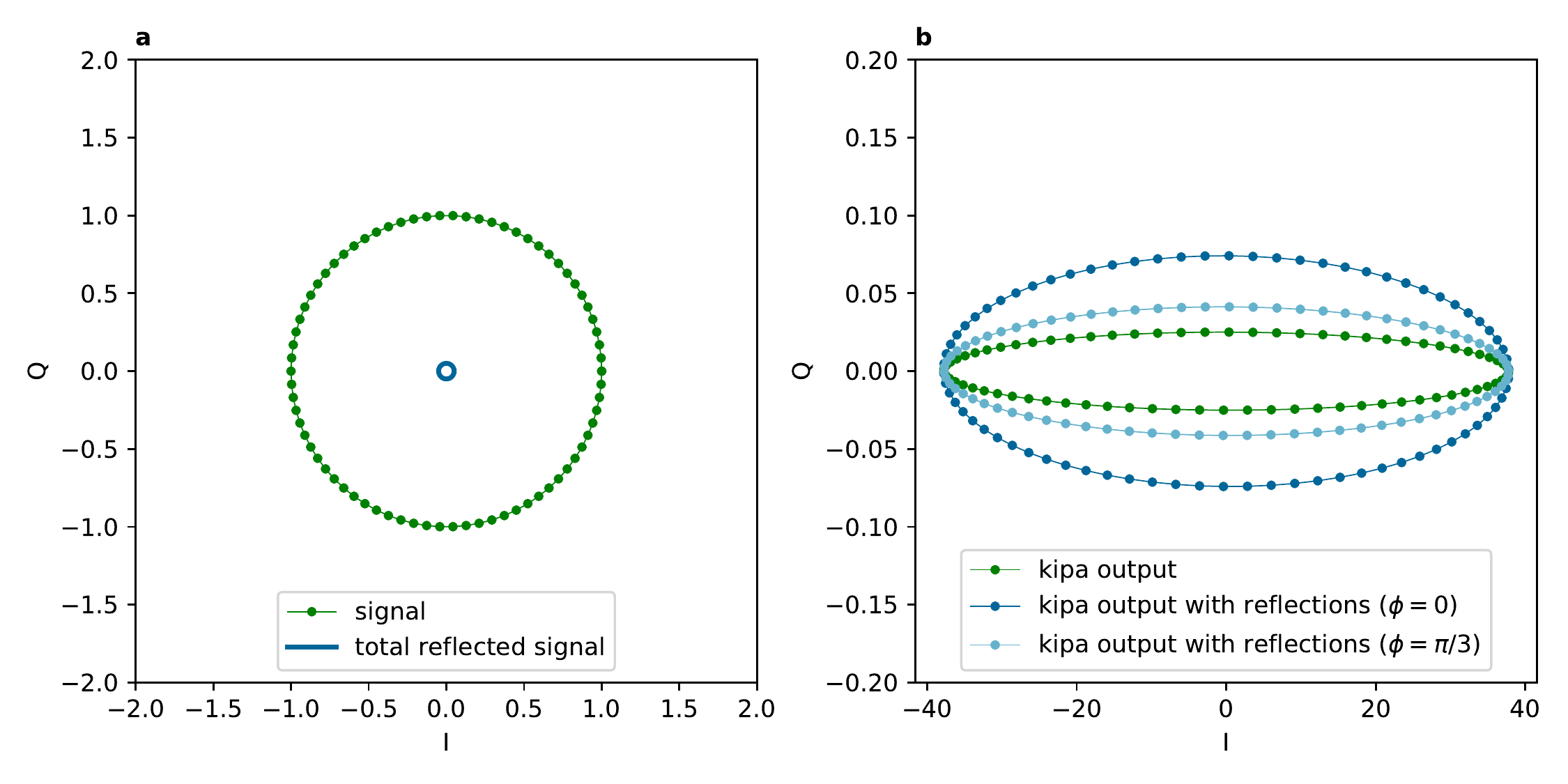}
    \caption{\label{fig:reflections_w_sqz}
        (a) An input signal (green) represented by points of constant amplitude on the IQ-plane, with a 5\% reflected signal (blue).
        (b) The theoretical output of the KIPA (green) superimposed with the total reflected signal with a phase shift of zero (dark blue) and a phase shift of $\pi/3$ (light blue). 
    }
\end{figure}

To be precise, we define `reflections' to be the total microwave signal that propagates towards the HEMT input that has not been amplified by the KIPA. The total reflected signal will have a constant amplitude that is a fraction of the input amplitude, and, relative to the KIPA output, will be offset in phase according to the difference in path length. Taking the vector sum between the total reflected signal and the phase sensitive output of the KIPA gives a resulting ellipse that we observe at the output of our fridge (see Fig.~S\ref{fig:reflections_w_sqz}). That is,
\begin{equation}
    \begin{pmatrix}
        \langle I_\text{out} \rangle \\
        \langle Q_\text{out} \rangle
    \end{pmatrix}
    = \Bigg[\mathcal{T}A_G(\varphi_p) + \mathcal{R} R(\phi)\Bigg]
    \begin{pmatrix}
        \langle I_\text{in} \rangle \\
        \langle Q_\text{in} \rangle
    \end{pmatrix} \label{eqn:sqz_w_reflection}
\end{equation}
where $\mathcal{T}$ is the coefficient of the input signal transmitted to the KIPA, $\mathcal{R}$ is reflection coefficient (with $\mathcal{T}^2 + \mathcal{R}^2 = 1$) and $R(\phi)$ is the standard rotation matrix that accounts for a phase shift of $\phi$. Fig.~S\ref{fig:reflections_w_sqz}b illustrates the effect of a 5\% reflection on the measurement of the output of a DPA. In the worst case of $\phi = 0$, the output of the KIPA and the reflected signal constructively interfere and degrade the observed deamplification by $\sim 9$ dB. The error introduced by the reflected signal will depend on the phase relationship between the KIPA output and the reflected signal, which in general is unknown. To proceed with the analysis, we define an in-phase reflection coefficient $\mathcal{R}'$ and set $\phi=0$. The in-phase reflection coefficient $\mathcal{R}'$ therefore represents a lower bound for the reflections in the setup needed to explain a given reduction in the observed deamplification level (see Fig.~3c of the main text). 

\begin{figure}
    \centering
    \includegraphics[width=\textwidth]{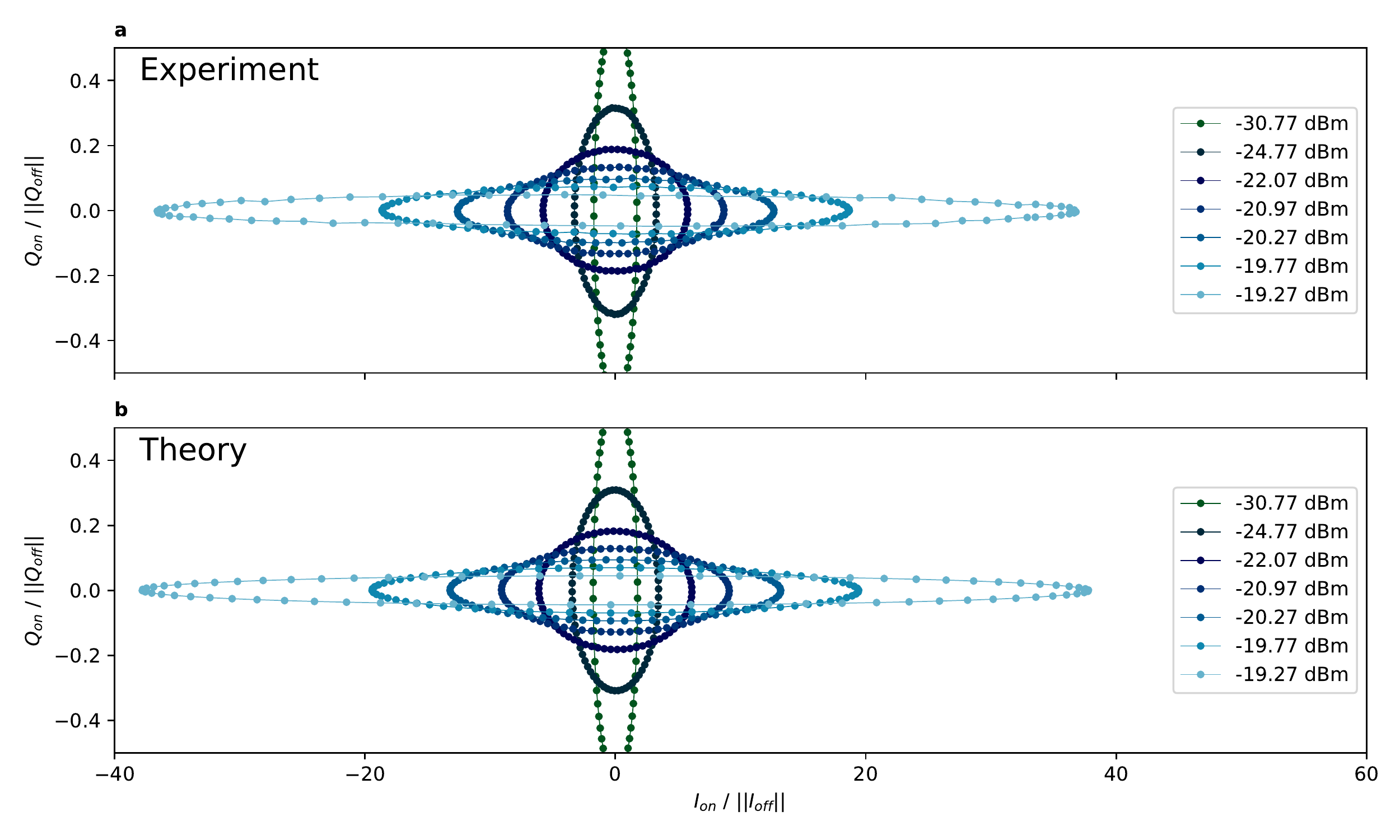}
    \caption{\label{fig:fitted_ellipses}
        (a) Ellipses measured by sweeping the phase of a fixed amplitude input, normalised by the amplitude of the input ($||I_\text{off}||$ or $|| Q_\text{off} ||$). This is the same data as is presented in Fig.~3b in the main text. Pump powers are shown in the legend. (b) Predicted ellipses from the DPA squeezing transformation with a 2\% in-phase reflection coefficient.
    }
\end{figure}

Using \cref{eqn:sqz_w_reflection} combined with the DPA parameters extracted from the fits in \cref{sec:phase_insens_gain} ($\kappa$, $|\xi|$, $\Delta$, etc.), we are able to predict the transformation of a unit magnitude input by the KIPA for different levels of in-phase reflection. We fit the in-phase reflection coefficient to be $\sim 2\%$ and find excellent agreement between theory and experiment (see Fig.~S\ref{fig:fitted_ellipses}).

The most likely sources of reflections from our setup are the connection from the coaxial lines to our bespoke PCB, and at the wire bonds between the PCB and the chip. Assuming $50~\Omega$ lines down to the sample, a $2\%$ reflection corresponds to an equivalent PCB impedance of:
\begin{equation}
    Z_\text{PCB} = Z_\text{CPW} \frac{1 - \mathcal{R}}{1 + \mathcal{R}} = 48~\Omega
\end{equation}
which is realistic accounting for the uncertainty in the design and manufacturing tolerances of the PCB and the temperature dependence of the materials.

The maximum deamplification level $G_S$ is defined as the greatest reduction in amplitude of a coherent input by the squeezing transformation, whilst $G_A$ is the corresponding increase in gain that occurs orthogonal to the axis of deamplification. $G_S$ and $G_A$ are extracted from the ellipse data of Fig.~3b and plot in Fig.~3c in the main text. We reproduce the ellipse measurement data here in Fig.~S\ref{fig:fitted_ellipses}, along with a set of ellipses generated using our theoretical model. We observe some asymmetry $G_S \neq G_A$ in the data, which is captured accurately by our model that includes weak reflections in the experimental setup (solid lines in in Fig.~3c). The ideal amplifier symmetrically transforms both quadratures (i.e. $G_S = G_A$) \cite{caves_1982}, however, according to our model for the squeezing transform, symmetry can also be broken if either $Q_i < \infty$ or $|\Delta| > 0$. While some asymmetry is expected, for our estimate of $Q_i = 10^5$, this asymmetry is small as is evident in Fig.~3c (dashed line) where we show the predicted $G_S$ for the reflection-less DPA measurement.

\section{Noise Squeezing Theory} \label{sec:sqz_theory}
In \cref{sec:sqz_tform_theory} we analyzed the gain of the KIPA in phase sensitive mode when coherent states were applied to its input. In this section we consider the case of a vacuum input state (i.e. noise) and derive expressions for the squeezing (noise deamplification) and anti-squeezing (noise amplification) properties of the KIPA. We assume that noise squeezing is measured over a narrow-band such that the frequency dependence of $g_s$ and $g_i$ may be ignored, allowing us to draw on the theory presented in \cref{sec:sqz_tform_theory}. In terms of the matrix elements $g_{ij}$ of the squeezing transformation matrix $A_G$, \cref{eqn:io_matrix} becomes:
\begin{align}
    \begin{pmatrix}
        I_\text{out} \\
        Q_\text{out}
    \end{pmatrix}
    &= A_G(\theta)
        \begin{pmatrix}
            I_\text{in} \\
            Q_\text{in}
        \end{pmatrix}
    + \sqrt{\frac{\gamma}{\kappa}} \big( A_G(\theta) + 1\big)
    \begin{pmatrix}
        I_{b} \\
        Q_{b}
    \end{pmatrix} \\
    &= \begin{pmatrix}
        g_{11} & g_{12} \\
        g_{21} & g_{22}
    \end{pmatrix}
    \begin{pmatrix}
        I_{in} \\
        Q_{in}
    \end{pmatrix}
    + \sqrt{\frac{\gamma}{\kappa}} \begin{pmatrix}
        g_{11} + 1 & g_{12} \\
        g_{21} & g_{22} + 1
    \end{pmatrix}
    \begin{pmatrix}
        I_{b} \\
        Q_{b}
    \end{pmatrix}
\end{align}

To obtain expressions for the vacuum squeezing, we model the input field as a vacuum state with variances $\langle \Delta I^2 \rangle = \langle \Delta Q^2 \rangle$ and zero mean: $\langle I \rangle = \langle Q \rangle = 0$. Assuming the bath and the input fields are uncorrelated, and using the fact that $\langle IQ \rangle + \langle QI \rangle = 0$, we may write a system of linear equations for the second order moments of the output quadratures:
\begin{equation}
    \begin{pmatrix}
        \langle I_\text{out}^2 \rangle \\
        \langle Q_\text{out}^2 \rangle
    \end{pmatrix}
    = \begin{pmatrix}
        g_{11}^2 & g_{12}^2 \\
        g_{21}^2 & g_{22}^2
    \end{pmatrix}
    \begin{pmatrix}
        \langle I_\text{in}^2 \rangle\\
        \langle Q_\text{in}^2 \rangle
    \end{pmatrix} + \frac{\gamma}{\kappa} \begin{pmatrix}
        (g_{11} + 1)^2 & g_{12}^2 \\
        g_{21}^2 & (g_{22} + 1)^2
    \end{pmatrix}
    \begin{pmatrix}
        \langle I_{b}^2 \rangle \\
        \langle Q_{b}^2 \rangle
    \end{pmatrix}
\end{equation}

Since the vacuum and bath fields are at the same temperature, we define $\langle \Delta I_v^2 \rangle = \langle I_\text{in}^2 \rangle = \langle I_b^2 \rangle = 1/4$ and $\langle \Delta Q_v^2 \rangle = \langle Q_\text{in}^2 \rangle = \langle Q_b^2 \rangle = 1/4$. Thus, the variances of the output quadratures are given by:

\begin{equation}
    \begin{pmatrix}
        \langle \Delta I_\text{out}^2 \rangle \\
        \langle \Delta Q_\text{out}^2 \rangle 
    \end{pmatrix}
    = \Bigg[ \begin{pmatrix}
        g_{11}^2 & g_{12}^2 \\
        g_{21}^2 & g_{22}^2
    \end{pmatrix}
    + \frac{\gamma}{\kappa} \begin{pmatrix}
        (g_{11} + 1)^2 & g_{12}^2 \\
        g_{21}^2 & (g_{22} + 1)^2
    \end{pmatrix} \Bigg]
    \begin{pmatrix}
        \langle \Delta I_{v}^2 \rangle \\
        \langle \Delta Q_{v}^2 \rangle 
    \end{pmatrix}
\end{equation}

The increase/decrease in quadrature variance as a function of the pump phase $\varphi_p$ is described by:
\begin{align}
    \mathcal{S}(\varphi_p) = 10\log_{10}\frac{\langle \Delta I_\text{out}^2 \rangle}{\langle \Delta I_v^2 \rangle}
                        = 10\log_{10}\bigg(g_{11}^2 + g_{12}^2 + \frac{\gamma}{\kappa}((g_{11} + 1)^2 + g_{12}^2)\bigg)
    \label{eqn:noise_variance_gain}
\end{align}
and the vacuum squeezing level $\mathcal{S}_v$ is defined by the minimum of $\mathcal{S}$:
\begin{equation}
    \mathcal{S}_v = \min_{\varphi_p} \mathcal{S}(\varphi_p)
\end{equation}

\begin{figure}
    \centering
    \includegraphics[width=\textwidth]{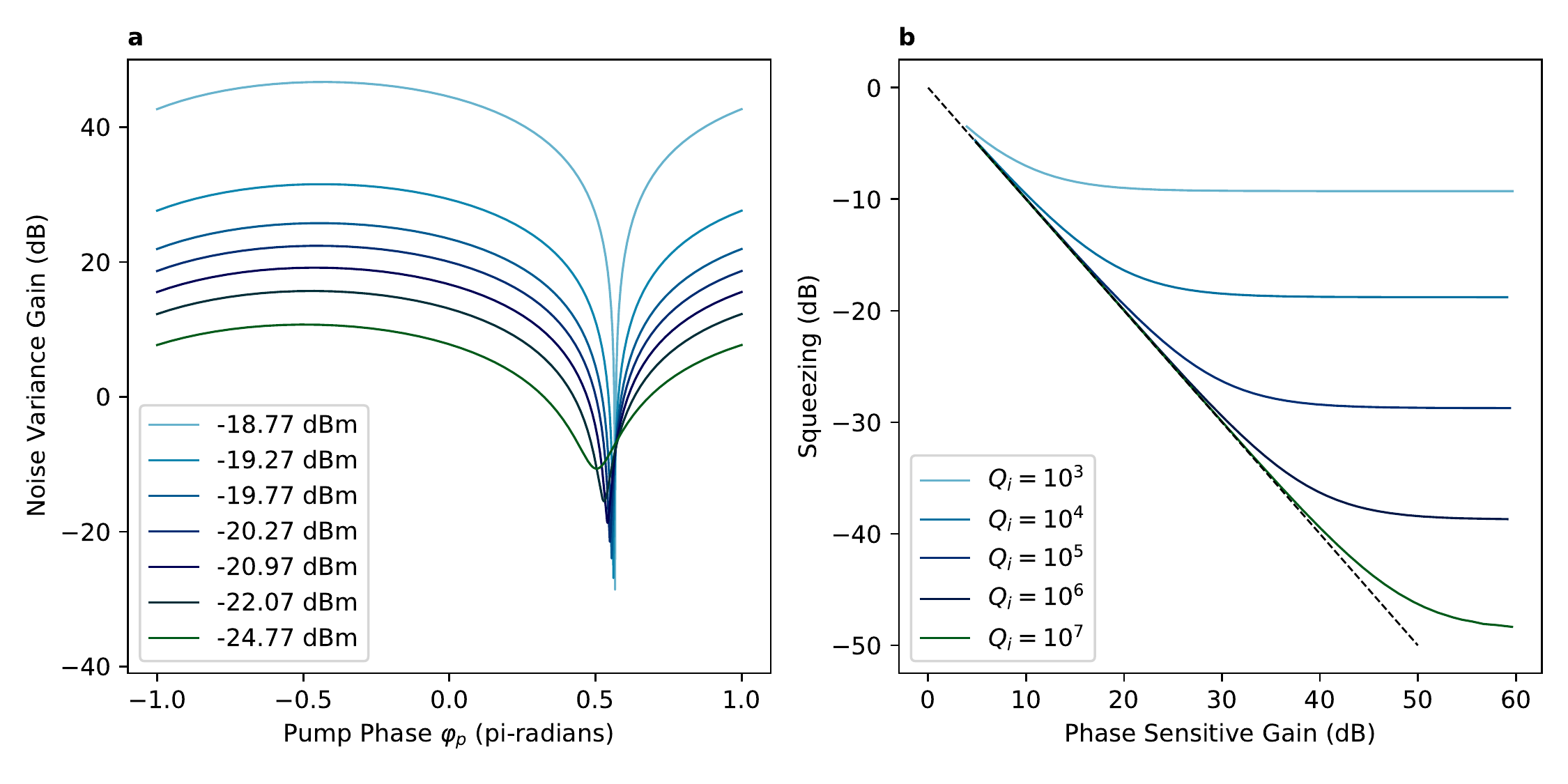}
    \caption{\label{fig:theoretical_squeezing}
        (a) The simulated noise variance gain $\mathcal{S}(\varphi_p)$ as a function of the pump phase using the DPA parameters obtained in \cref{sec:phase_insens_gain}, for a range of pump powers. We fix the internal quality factor to $Q_i = 10^5$ in this calculation.
        (b) The vacuum squeezing level $\mathcal{S}_v$, as a function of the phase sensitive gain using the DPA parameters obtained in \cref{sec:phase_insens_gain} for different internal quality factors $Q_i$. The line of symmetric phase sensitive gain is plotted (black dashed line).
    }
\end{figure}

Using the DPA parameters extracted from the phase-insensitive gain features (see \cref{sec:phase_insens_gain}), we can simulate the expected noise variance gain (\cref{eqn:noise_variance_gain}) as a function of the pump phase. The results are depicted in Fig.~S\ref{fig:theoretical_squeezing}a, where we observe a similar phase dependent response as was measured in \cref{sec:phase_sens_gain} for strong coherent inputs. In contrast to the phase-sensitive gain for coherent inputs, the coupling of the bath mode into the cavity requires a strictly asymmetric noise variance gain such that $(\min_{\varphi_p} S(\varphi_p) \times \max_{\varphi_p} S(\varphi_p)) \ge 1$, where equality holds only in the limit of $Q_i \to \infty$. We observe a weak shift in the pump phase corresponding to the point of maximum squeezing as the pump power increases, which is a consequence of the non-zero detuning between the cavity and the pump $\Delta$.

Equipped with this squeezing model and a realistic set of resonator parameters, we can study the effect of $Q_i$ on the maximum attainable squeezing. Fig.~S\ref{fig:theoretical_squeezing}b plots the vacuum squeezing level $\mathcal{S}_v$ against the maximum variance gain, or anti-squeezing gain. In the limit of no losses, Cave's theory predicts symmetric squeezing and anti-squeezing with zero noise photons contributed by the amplifier \cite{caves_1982}. We observe here that the squeezing/anti-squeezing relationship of the KIPA closely follows the expected symmetric behaviour before the squeezing level plateaus to a constant level as the anti-squeezing gain increases. The squeezing level plateaus as the total cavity fluctuations are limited by the bath mode variance, which is not squeezed by the KIPA since:
\begin{equation}
    \bigg( A_G + 1\bigg)
    \begin{pmatrix}
        \text{Var}(I_b) \\
        \text{Var}(Q_b)
    \end{pmatrix}
    \ge 
    \begin{pmatrix}
        \text{Var}(I_b) \\
        \text{Var}(Q_b)
    \end{pmatrix}
\end{equation}

We observe an approximate $10$~dB improvement in the maximum achievable squeezing for each order of magnitude increase in $Q_i$. The order of magnitude improvement in squeezing performance is a result of the corresponding order of magnitude decrease in $\gamma/\kappa$, which sets the magnitude of the bath variance contribution to the KIPA output (see \cref{eqn:io_matrix}). For a $Q_i = 10^5$, our theory predicts up to $\mathcal{S}_v \approx -29$~dB of squeezing could be produced by the KIPA, corresponding to approximately $40$ dB of phase sensitive gain.

\section{Noise Temperature} \label{sec:noise_temp_theory}
\subsection{Non-Degenerate Noise Temperature Theory} \label{sec:nd_noise_temp}
The output fluctuations of the KIPA operating as a non-degenerate amplifier are found from \cref{eqn:boutin_io}) to be:
\begin{equation}
\begin{aligned}
    \langle \Delta I_\text{out}^2 \rangle &= \left< \left[\frac{1}{2}\left(a^\dagger_\text{out} + a_\text{out}\right)\right]^2\right>\\
        &= \bigg(|g_s|^2 + \frac{\gamma}{\kappa}|g_s + 1|^2\bigg)\bigg(\frac{n_\text{th}}{2} + \frac{1}{4}\bigg) + |g_i|^2\bigg(1 + \frac{\gamma}{\kappa}\bigg)\bigg(\frac{n_\text{th}}{2} + \frac{1}{4}\bigg)
\end{aligned}
\end{equation}
where the signal and idler gains ($g_s(\omega)$ and $g_i(\omega)$) depend on the frequency of the signal being amplified. Here we assume that the input and bath fields (both signal and idler modes) have a thermal occupation $\langle a_\text{in}^\dagger a_\text{in}\rangle = \langle b_\text{in}^\dagger b_\text{in}\rangle = n_\text{th}$.

One useful identity of the DPA is the relationship between the signal and idler gains \cite{boutin_2017}:
\begin{equation}
    |g_i|^2\bigg(1 + \frac{\gamma}{\kappa}\bigg) = |g_s|^2 + \frac{\gamma}{\kappa}|g_s + 1|^2 - 1 \label{eqn:dpa_commutation_identiy}
\end{equation}
which holds for all $\omega$ (see \cref{eqn:boutin_refltion_params}), and is a by-product of the KIPA output field satisfying the commutation relation $[a_\text{out}, a_\text{out}^\dagger] = 1$. Substituting \cref{eqn:dpa_commutation_identiy} into our expression for the quadrature fluctuations along $I$, we obtain:
\begin{align}
    \langle \Delta I_\text{out}^2 \rangle 
        = \bigg(|g_s|^2 + \frac{\gamma}{\kappa}|g_s + 1|^2\bigg)\bigg(n_\text{th} + \frac{1}{2}\bigg) - \frac{n_\text{th}}{2} - \frac{1}{4} \label{eqn:output_fluc_partial}
\end{align}
Referring the quadrature fluctuations to the input of the KIPA and subtracting the vacuum contribution, we find:
\begin{align}
    \frac{\langle \Delta I_\text{out}^2 \rangle}{|g_s|^2} - \frac{1}{4} 
        &= \bigg(1 + \frac{\gamma}{\kappa}\frac{|g_s + 1|^2}{|g_s|^2}\bigg)\bigg(n_\text{th} + \frac{1}{2}\bigg) - \frac{n_\text{th}}{2|g_s|^2} - \frac{1}{4|g_s|^2} - \frac{1}{4}\\
        &\ge \frac{1}{4}\bigg(1 - \frac{1}{|g_s|^2}\bigg) + \frac{\gamma}{\kappa}\frac{|g_s + 1|^2}{2|g_s|^2}\\
        &\ge \frac{1}{4}\bigg(1 - \frac{1}{|g_s|^2}\bigg)
\end{align}
where in the second line we assume zero temperature ($n_\text{th} = 0$) and in the third line we assume no loss ($\gamma = 0$). As required by Cave's fundamental theorem of phase sensitive amplifiers, the KIPA/DPA adds $1/4$ photons to the input referred noise in the limit of high gain \cite{caves_1982}. Equality only holds in the limit of zero temperature and no losses.

\begin{figure}
    \includegraphics[width=\textwidth]{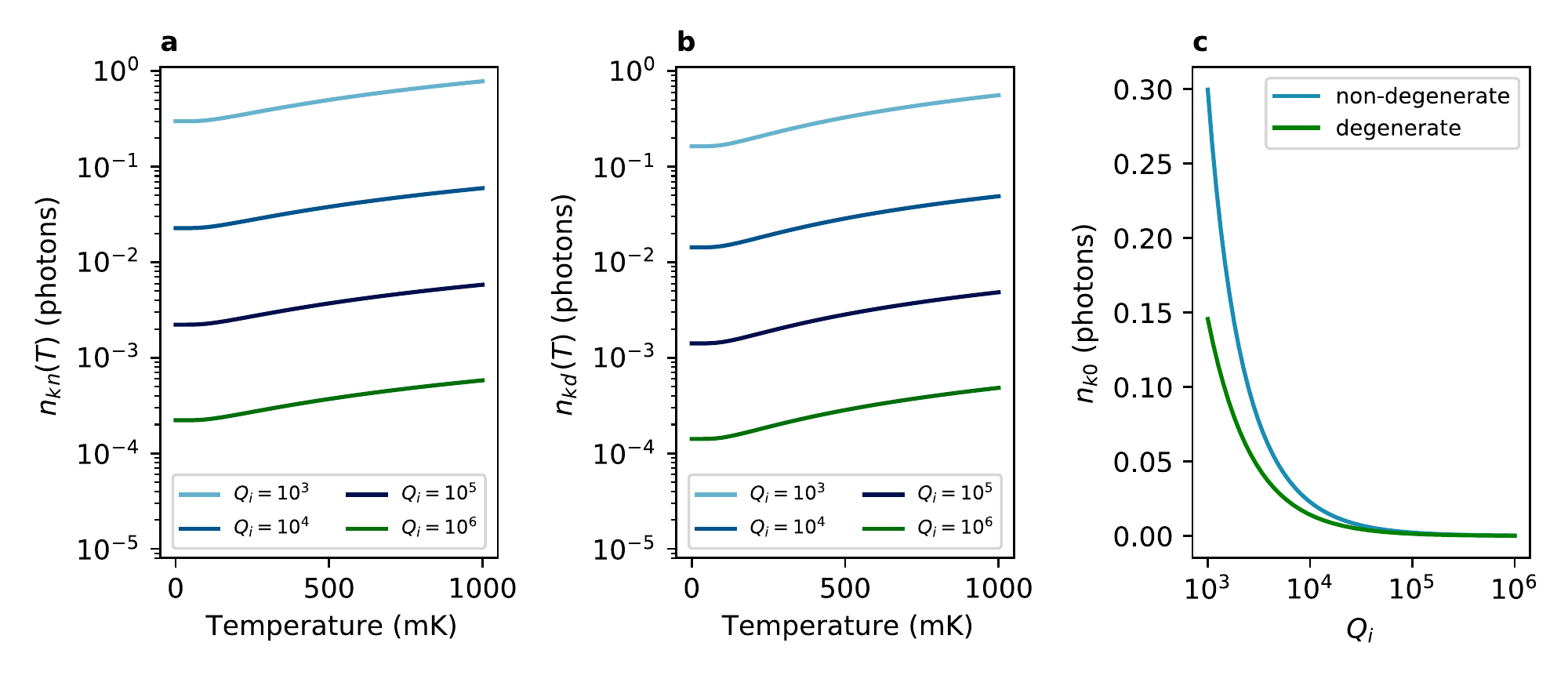}
    \caption{\label{fig:noise_temp_theory}
        (a) The simulated non-degenerate noise photon number $n_{kn}$ as a function of temperature for different values of $Q_i$.
        (b) The simulated degenerate noise photon number $n_{kd}$ as a function of temperature for different values of $Q_i$.
        (c) The simulated non-degenerate (degenerate) noise photon number at zero temperature as a function of the internal quality factor $Q_i$.
        All calculations performed at a non-degenerate (degenerate) gain of $G_k = 25$ dB ($31$ dB).
    }
\end{figure}

To maintain consistency with the input-output models for phase sensitive amplifiers used later in this section (e.g. see \cref{eqn:amplifier_eqn}), we write the phase insensitive output of the KIPA as follows:
\begin{equation}
    \langle \Delta I_\text{out}^2 \rangle = G_k \bigg(\frac{n_\text{th}}{2} + \frac{1}{4}\bigg) + (G_k - 1)\bigg(\frac{n_\text{th}}{2} + \frac{n_{kn}}{2} + \frac{1}{4}\bigg) \label{eqn:nt_kipa_out_nd}
\end{equation}
where $G_k = |g_s|^2$ and $n_{kn}$ is the input referred noise contribution of the KIPA in non-degenerate mode. Comparing \cref{eqn:output_fluc_partial} with the $(G_k - 1)$ term from this expression we obtain a relation for the additional noise photons contributed by the KIPA $n_{kn}$:
\begin{equation}
\begin{aligned}
    n_{kn} 
        &= \frac{2}{|g_s|^2 - 1}\bigg(\langle \Delta I_\text{out}^2 \rangle - |g_s|^2\bigg(\frac{n_\text{th}}{2} + \frac{1}{4}\bigg)\bigg) - n_\text{th} - \frac{1}{2} \\
        &= \frac{2}{|g_s|^2 - 1}\bigg(\bigg(\frac{n_\text{th}}{2} + \frac{1}{4}\bigg)(|g_s|^2 - 1) + \frac{\gamma}{\kappa}|g_s + 1|^2\bigg(n_\text{th} + \frac{1}{2}\bigg)\bigg) - n_\text{th} - \frac{1}{2}\\
        &= \frac{\gamma}{\kappa}\frac{|g_s + 1|^2}{|g_s|^2 - 1}(2n_\text{th} + 1)
\end{aligned}
\end{equation}
The temperature dependence for $n_{kn}$ is depicted in Fig.~S\ref{fig:noise_temp_theory} for various internal quality factors, and using the same DPA parameters as were measured previously. Compared to $n_\text{th}$ the change in $n_{kn}$ is small across the range of internal quality factors considered. At zero temperature $n_{kn}$ appears to decrease by an approximate order of magnitude for every increase in the order of magnitude for $Q_i$, further motivating the desire to maximise the $Q_i$ of a DPA.

Operating the KIPA such that $\hbar \omega_0 \ll k_B T$, we have:
\begin{equation}
    n_{kn0} = \frac{\gamma}{\kappa}\frac{|g_s + 1|^2}{|g_s|^2 - 1}
\end{equation}
We plot $n_{kn0}$ as a function of $Q_i = \omega_0/\gamma$ and observe rapid convergence to zero as $Q_i \to \infty$.

\subsection{Degenerate Noise Temperature Theory}
The output fluctuations of the KIPA in degenerate mode as a function of the pump phase $\varphi_p$ are given by (see \cref{sec:sqz_theory}):
\begin{equation}
    \langle \Delta I_\text{out}(\varphi_p)^2 \rangle
        = |g_s + g_i^*(\varphi_p)|^2 \bigg(\frac{n_\text{th}}{2} + \frac{1}{4}\bigg) + \frac{\gamma}{\kappa}|g_s + 1 + g_i^*(\varphi_p)|^2\bigg(\frac{n_\text{th}}{2} + \frac{1}{4}\bigg)
\end{equation}
with phase sensitive power gain $G_k(\varphi_p) = |g_s + g_i^*(\varphi_p)|^2$. Referred to the KIPA input, the excess quadrature fluctuations contributed by the amplifier are:
\begin{equation}
    \frac{\langle \Delta I_\text{out}(\varphi_p)^2 \rangle}{G_k(\varphi_p)} - \frac{1}{4} 
        = \frac{n_\text{th}}{2} + \frac{\gamma}{\kappa} \frac{|g_s + 1 + g_i^*(\varphi_p)|^2}{|g_s + g_i^*(\varphi_p)|^2}\bigg(\frac{n_\text{th}}{2} + \frac{1}{4}\bigg) \ge 0
\end{equation}

As predicted by Caves, the excess quadrature fluctuations referred to the input can be as small as zero in limit of $\hbar \omega \ll k_B T$ and provided there are no losses in the system (i.e. $\gamma = 0$) \cite{caves_1982}.

Writing $G_k = |g_s + g_i^*(\varphi_p)|^2$, we define a similar expression to \cref{eqn:nt_kipa_out_nd} for the phase sensitive amplifier along the amplified quadrature:
\begin{equation}
    \langle \Delta I_\text{out}^2 \rangle
        = G_k \bigg(\frac{n_\text{th}}{2} + \frac{1}{4}\bigg) + (G_k - 1)\bigg(\frac{n_{kd}}{2}\bigg) \label{eqn:nt_kipa_out_d}
\end{equation}
with,
\begin{align}
    n_{kd} &= \frac{\gamma}{\kappa}\frac{|g_s + 1 + g_i^*|^2}{|g_s + g_i^*|^2 - 1}\bigg(n_\text{th} + \frac{1}{2}\bigg)\label{eq:nk}\\
    n_{kd0} &= \frac{\gamma}{2\kappa}\frac{|g_s + 1 + g_i^*|^2}{|g_s + g_i^*|^2 - 1}
\end{align}

Again, we simulate $n_{kd}$ for varied internal quality factors and temperatures and find similar behaviour to the non-degenerate case. In the limit of high gain, the minimum noise added $n_{kd0}$ is approximately half the corresponding noise added in the non-degenerate case (see Fig.~S\ref{fig:noise_temp_theory}).

\subsection{Noise Temperature Measurement}
\begin{figure}
    \centering
    \includegraphics[width=\textwidth]{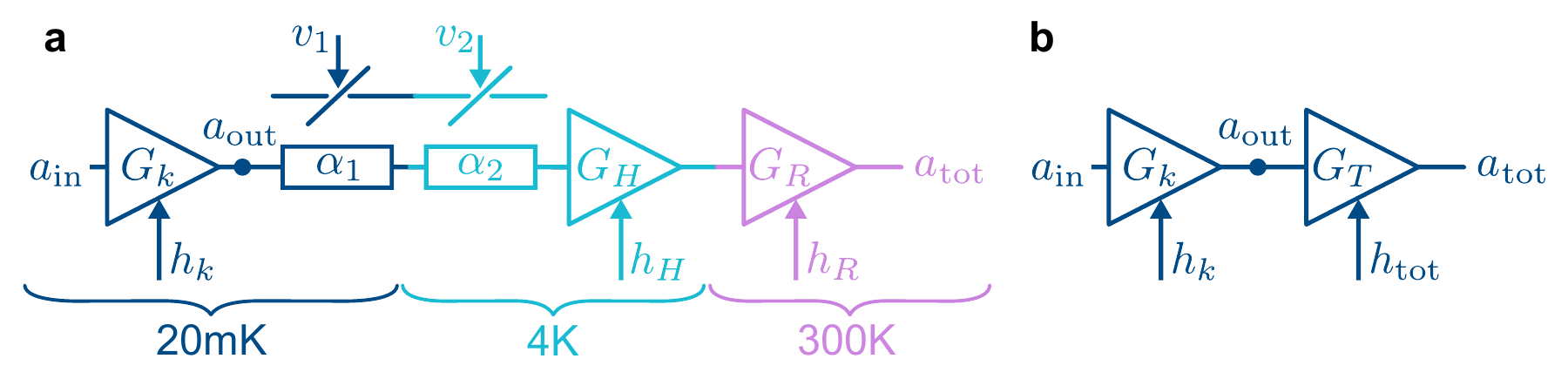}
    \caption{\label{fig:noise_temp_model}
        (a) The complete detection chain model, consisting of the KIPA and attenuator $\alpha_1$ at 20~mK, attenuator $\alpha_2$ and the HEMT at 4~K, and the room temperature amplifier at 300~K. The attenuators are modeled as beam splitters, mixing in the thermal operators $v_1$ and $v_2$ with the detected field as it propagates along the detection chain. Each amplifier contributes noise to its output, denoted here by the field operators $h_{k}$ (KIPA), $h_H$ (HEMT), and $h_R$ (room temperature amplifier).
        (b) The simplified detection chain model, where the attenuators and amplifiers after the KIPA may be modeled as an effective amplifier with gain $G_T$ and noise field $h_\text{tot}$.
    }
\end{figure}

The detection chain of the KIPA consists of a series of amplifiers and attenuators, which we depict in Fig.~S\ref{fig:noise_temp_model}a. Directly after the KIPA there are microwave losses associated with the diplexer, the circulator and the microwave lines. To model the detection chain we divide these losses into two effective attenuators, one at 20~mK and the other at 4~K. Each attenuator acts like an optical beam splitter, where the transmitted field is reduced by $\sqrt{\alpha}$ and the open port mixes the thermal field $v$ into the output according to the beam splitter equation \cite{walls_2008}:
\begin{equation}
    a_\text{out} = \sqrt{\alpha} a_\text{in} + \sqrt{1 - \alpha}v \label{eqn:beam_splitter}
\end{equation}

At 4~K we have the HEMT amplifier, followed by a second microwave amplifier at room temperature. Each amplifier contributes additional noise to its output field \cite{caves_1982}: 
\begin{equation}
    a_\text{out} = \sqrt{G_\text{amp}} a_\text{in} + \sqrt{G_\text{amp} - 1}h^\dagger \label{eqn:amplifier_eqn}
\end{equation}

Combining the attenuator models for $\alpha_1$ and $\alpha_2$ (\cref{eqn:beam_splitter}), with the amplifier models for the HEMT and room temperature amplifier (\cref{eqn:amplifier_eqn}), we may simplify the detection chain to a single equivalent amplifier with gain $G_T$ and noise contribution $h_\text{tot}$ (see Fig.~S\ref{fig:noise_temp_model}b). The total output field at the end of the detection chain is given by:
\begin{equation}
    a_\text{tot} = \sqrt{G_T}a_\text{out} + \sqrt{G_T - 1}h^\dagger_\text{tot}
\end{equation}
where,
\begin{align}
    G_T &= G_R G_H \alpha_1 \alpha_2\\
    h^\dagger_\text{tot} &= \sqrt{\frac{G_R G_H}{G_T - 1}}\Bigg[
        \sqrt{\alpha_1(1 - \alpha_1)}v_1 +
        \sqrt{(1 - \alpha_2)}v_2 +
        \sqrt{\frac{G_H - 1}{G_H}} h^\dagger_H +
        \sqrt{\frac{G_R - 1}{G_R G_H}} h^\dagger_R
    \Bigg]
\end{align}

Rewriting the output field $a_\text{out}$ as a pump phase dependent quadrature operator $I_\text{out}(\varphi_p) = (a_\text{out}^\dagger e^{-i\varphi_p} + a_\text{out} e^{i\varphi_p})/2$, we have:
\begin{equation}
    I_\text{tot}(\varphi_p) = \sqrt{G_T} I_\text{out}(\varphi_p) + \sqrt{G_T - 1}I_h(-\varphi_p)
\end{equation}
where $I_\text{out}(\varphi_p)$ is the pump phase dependent quadrature operator at the KIPA output, and $I_h$ is the detection chain noise quadrature operator $I_h(-\varphi_p) = (h_\text{tot}^\dagger e^{i\varphi_p} + h_\text{tot} e^{-i\varphi_p})/2$.

Assuming $h_\text{tot}$ and $a_\text{out}$ are composed of uncorrelated thermal states the quadrature fluctuations at the detector simplify to:
\begin{align}
    \langle \Delta I_\text{tot}^2 \rangle 
        &= G_T \langle \Delta I_\text{out}^2 \rangle + (G_T - 1) \langle \Delta I_h^2 \rangle\\
        &= G_T \langle \Delta I_\text{out}^2 \rangle + (G_T - 1) \bigg(\frac{n_\text{sys}}{2} + \frac{1}{4}\bigg)
\end{align}
where we introduce the effective system noise photon number $n_\text{sys}$:
\begin{align} 
    n_\text{sys}
        &= \langle h_\text{tot}^\dagger h_\text{tot} \rangle\\
        &= \frac{G_R G_H}{G_T - 1}\bigg[\alpha_2(1 - \alpha_1)(n_\text{20mK}+ 1) + (1 - \alpha_2)(n_\text{4K} + 1) + \frac{G_H - 1}{G_H}n_H + \frac{G_R - 1}{G_R G_H}n_R\bigg]
    \label{eqn:nsys}
\end{align}

The average microwave noise power that would be measured by a spectrum analyzer is simply the sum of the output quadrature variances:
\begin{equation}
    P_\text{tot} = z(\langle \Delta I_\text{tot}^2 \rangle + \langle \Delta Q_\text{tot}^2 \rangle)
\end{equation}
We introduce the parameter $z$ here that converts the units from photons to Watts as is measured by the spectrum analyzer over a certain measurement bandwidth resolution.

In non-degenerate operation, the variance of both the KIPA output and the system noise fields are independent of the pump phase, allowing us to write the measured microwave power as: 
\begin{align}
    P_\text{tn} 
        &= zG_T (\langle \Delta I_\text{out}^2 \rangle + \langle \Delta Q_\text{out}^2 \rangle) + z(G_T - 1)\bigg(n_\text{sys} + \frac{1}{2}\bigg) \\
        &= 2zG_T \langle \Delta I_\text{out}^2 \rangle + z(G_T - 1)\bigg(n_\text{sys} + \frac{1}{2}\bigg) \label{eqn:avg_noise_power}
\end{align}

In non-degenerate mode, the output fluctuations of the KIPA are given by (see Section \ref{sec:nd_noise_temp}):
\begin{align}
    \langle \Delta I_\text{out}^2 \rangle 
        &= G_k \bigg(\frac{n_\text{th}}{2} + \frac{1}{4}\bigg) + (G_k - 1)\bigg(\frac{n_\text{th}}{2} + \frac{n_{kn}}{2} + \frac{1}{4}\bigg) \label{eqn:kipa_thermal_output}
\end{align} 
with $G_k = |g_s|^2$ as defined in \cref{eqn:boutin_refltion_params}, thermal noise population $n_\text{th} = \langle a^\dagger_\text{in} a_\text{in} \rangle$, and an additional number of noise photons added by the KIPA $n_{kn} = \langle h^\dagger_k h_k \rangle$. In the non-degenerate case, the idler mode contributes a minimum $n_\text{th}/2 + 1/4$ input-referred photons to the variance of each quadrature at the signal frequency, while an additional $n_{kn}/2$ photons arise from internal cavity losses. The excess noise $n_{kn}$ is expected to vary with temperature (see \cref{sec:nd_noise_temp}), however, for $Q_i > 10^4$ this dependence is negligible since $n_\text{th} \gg n_{kn}$ and therefore we approximate $n_{kn} \approx n_{kn0}$ to be constant with temperature.

Substituting \cref{eqn:kipa_thermal_output} into \cref{eqn:avg_noise_power}, we arrive at:
\begin{equation}
    P_\text{tn} 
        = zG_T G_k \bigg(n_\text{th} + \frac{1}{2}\bigg) + zG_T(G_k - 1)\bigg(n_\text{th} + n_{kn0} + \frac{1}{2}\bigg) + z(G_T - 1)\bigg(n_\text{sys} + \frac{1}{2}\bigg) \label{eqn:avg_noise_power_ii}
\end{equation}

Both $n_\text{sys}$ and the conversion factor $z G_T$ are unknown. We begin by finding $zG_T$, observing that when the KIPA is off (i.e. $G_k = 1$), \cref{eqn:avg_noise_power_ii} simplifies to:
\begin{equation}
    P_\text{off} = zG_T\bigg(n_\text{th} + \frac{1}{2}\bigg) + z(G_T - 1)\bigg(n_\text{sys} + \frac{1}{2}\bigg)
    \label{eqn:poff_nt}
\end{equation}

Evaluating the difference in power between when the KIPA is on compared to off removes the dependence on $n_\text{sys}$:
\begin{equation}
    P_\text{tn} - P_\text{off}
        = zG_T (G_k - 1) \bigg(2n_\text{th} + n_{kn0} + 1\bigg) 
    \label{eqn:pdiff_nt}
\end{equation}

\begin{figure}
    \includegraphics[width=\textwidth]{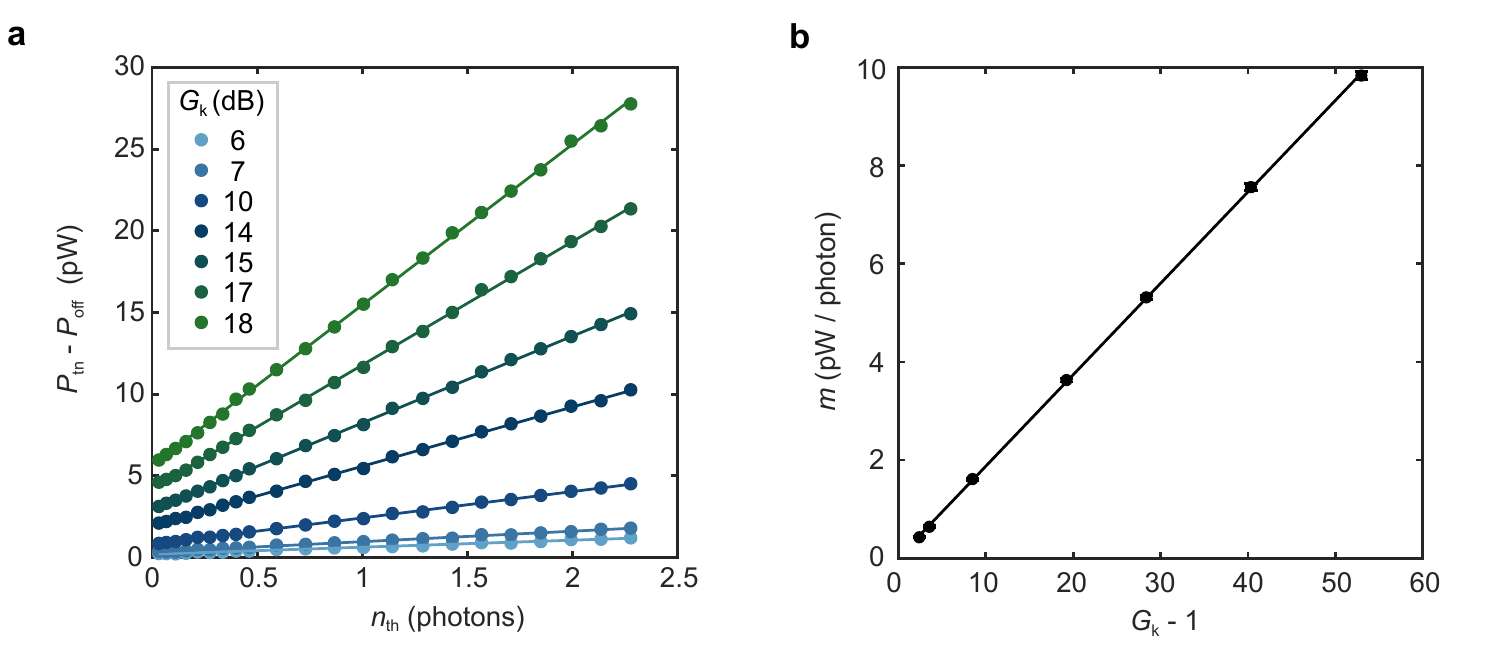}
    \caption{\label{fig:GT_measurement}
        (a) The difference power reported by the spectrum analyzer for the KIPA on vs off in non-degenerate mode as a function of the thermal photon population at the mixing plate for different non-degenerate (ND) gains (see legend). Solid lines are linear fits.
        (b) The gradient $m$ of the linear fits presented in (a) ($P_\text{on} - P_\text{off} = m\cdot n_\text{th} + b$) vs $G_k - 1 = |g_s|^2 - 1$.
    }
\end{figure}

To extract $zG_T$ we sweep the temperature of the mixing chamber of our dilution refrigerator (and thus $n_\text{th}$) while operating the KIPA as a non-degenerate amplifier ($500$~kHz detuned from $\omega_p/2$). At each temperature, we measure the noise power at the output of our detection chain using a spectrum analyzer configured in zero-span mode with a measurement bandwidth of $130$ kHz. We constrain the experiment to non-degenerate gains below $20$~dB, since below this the KIPA gain responses are completely flat over the 500~kHz detuned measurement band and we can therefore approximate $G_k$ by measuring the gain using a (narrow band) coherent tone. At each KIPA gain $G_k = |g_s|^2$, we expect the difference in power to increase linearly according to $P_\text{tn} - P_\text{off} = m\cdot n_\text{th} + b$ with gradient $m = 2zG_T(G_k - 1)$.  The data is shown in Fig.~S\ref{fig:GT_measurement}a, which displays a clear linearity with $n_\text{th}$ for various non-degenerate gains. We plot $m$ against $G_k - 1$ and extract the conversion factor $zG_T = 93.2(10)$ fW/photon (see Fig.~S\ref{fig:GT_measurement}b). Considering \cref{eq:nk}, we see that the $zG_T$ determined with this method may be smaller by a factor of approximately $\sim 1+ \gamma/\kappa$. Given our estimates of the loss in the KIPA (see \cref{sec:losses}), we believe this error is small (i.e. $< 5\%$). Assuming the lower-bound for $Q_i$, we evaluate the uncertainty in the conversion factor to be $zG_T = 93.2(53)$ fW/photon.

 Knowing $zG_T$, we may extract the noise temperature of the KIPA in non-degenerate mode by considering the noise referred to the input of the KIPA in photon units:
\begin{equation}
    \begin{aligned}
        n_\text{tn} &= \frac{P_\text{tn}}{zG_T G_k} \approx (2n_\text{th} + n_{kn0} + 1) + \frac{1}{G_k}\bigg(n_\text{sys} - n_\text{th} - n_{kn0}\bigg)\\
        &\approx (2n_\text{th} + n_{kn0} + 1) + \frac{n_\text{sys}}{G_k}\label{eqn:ntot}
    \end{aligned}
\end{equation}
where in the second line we assume that $n_\text{sys} \gg n_\text{th} + n_{kn0}$. We use this equation to fit the data in Fig.~4c of the main text and extract $n_\text{tn}^\infty = 2n_\text{th} + n_{kn0} + 1 = 1.18(9)$ photons and $n_\text{sys} = 80.0(46)$ photons, where $n_\text{tot}^\infty$ is the input-referred noise in the limit of infinite KIPA gain. To validate $n_\text{sys}$, we substitute data-sheet values for the HEMT and room temperature amplifier into \cref{eqn:nsys} and estimate $\alpha_1$ and $\alpha_2$ based on manufacturer values for cable, circulator and diplexer insertion losses. \cref{eqn:nsys} gives $n_\text{sys} \approx 64$ photons -- a reasonable agreement provided the uncertainty in the estimated losses.

We turn our attention now to the degenerate gain. Because the fluctuations along one quadrature of the KIPA output are squeezed and are therefore considerably smaller than the fluctuations along the orthogonal amplified quadrature, the total noise power measured at the spectrum analyzer may be approximated by:
\begin{align}
    P_\text{td} 
        &= zG_T (\langle \Delta I_\text{out}^2 \rangle + \langle \Delta Q_\text{out}^2 \rangle) + z(G_T - 1)\bigg(n_\text{sys} + \frac{1}{2}\bigg) \\
        &\approx zG_T \langle \Delta I_\text{out}^2 \rangle + z(G_T - 1)\bigg(n_\text{sys} + \frac{1}{2}\bigg) \label{eqn:avg_noise_power_degen}
\end{align}

From \cref{eqn:nt_kipa_out_d}, we have:
\begin{equation}
    P_\text{td} 
        = zG_T G_k \bigg(\frac{n_\text{th}}{2} + \frac{1}{4}\bigg) + zG_T(G_k - 1)\bigg(\frac{n_{kd0}}{2}\bigg) + z(G_T - 1)\bigg(n_\text{sys} + \frac{1}{2}\bigg)
\end{equation}
giving: 
\begin{equation}
\begin{aligned}
    n_\text{td} &= \frac{P_\text{td}}{zG_T G_k} \approx \frac{n_\text{th}}{2} + \frac{n_{kd0}}{2} + \frac{1}{4} + \frac{1}{G_k}\bigg(n_\text{sys} + \frac{1}{2} - \frac{n_{kd0}}{2}\bigg)\\
    &\approx \frac{1}{4}(2n_\text{th} + 2n_{kd0} + 1) + \frac{n_\text{sys}}{G_k}\label{eqn:ntot_ii}
\end{aligned}
\end{equation}

As before, we fit \cref{eqn:ntot_ii} to the measured noise power at the spectrum analyzer, referred to the input of the KIPA and expressed in photons. The results are depicted in Fig.~4c of the main text alongside the non-degenerate measurement. We find $n_\text{td}^\infty = (2n_\text{th} + 2n_{kd0} + 1)/4 = 0.31(5)$ photons, very close to the quantum-limited value of 0.25 photons.

We can set a bound on $n_{kd}$ ($n_{kn}$) by estimating the loss in our KIPA. The reflection magnitude response recorded with the KIPA off appears flat within the 0.7~dB measurement ripple, which implies $Q_i > 3,350$ and thus $n_{kd} < 0.03$ ($n_{kn} < 0.06$), see \cref{sec:losses} details. High internal quality factors are typical of planar NbTiN resonators, with $Q_i \sim 100,000$ often observed for our PBG resonators with higher external quality factors (\cref{sec:losses}).

\subsection{Signal Line Attenuation}
In Fig.~4b of the main text we plot the input-referred number of photons recorded in the presence of an applied coherent tone, with the KIPA in three different configurations: degenerate mode, non-degenerate mode and off. We calculate the input-referred number of photons by dividing the measured output power by $zG_TG_k$ and then the equivalent input-referred power by multiplying the number of photons by $\hbar\omega_0B$, where $B = 1$~kHz is the measurement bandwidth resolution. Knowing the power at the output of the signal generator ($-60$~dBm) used in this measurement, the input-referred coherent tone power ($-132$~dBm) can be used to calculate a 72~dB loss along the input signal line. This loss is consistent with the 60~dB of fixed attenuation in our setup, plus our estimates for additional cable and component insertion loss based on manufacturer data-sheets.

\section{KIPA Losses}\label{sec:losses}
\begin{figure}
    \centering
    \includegraphics[width=0.7\textwidth]{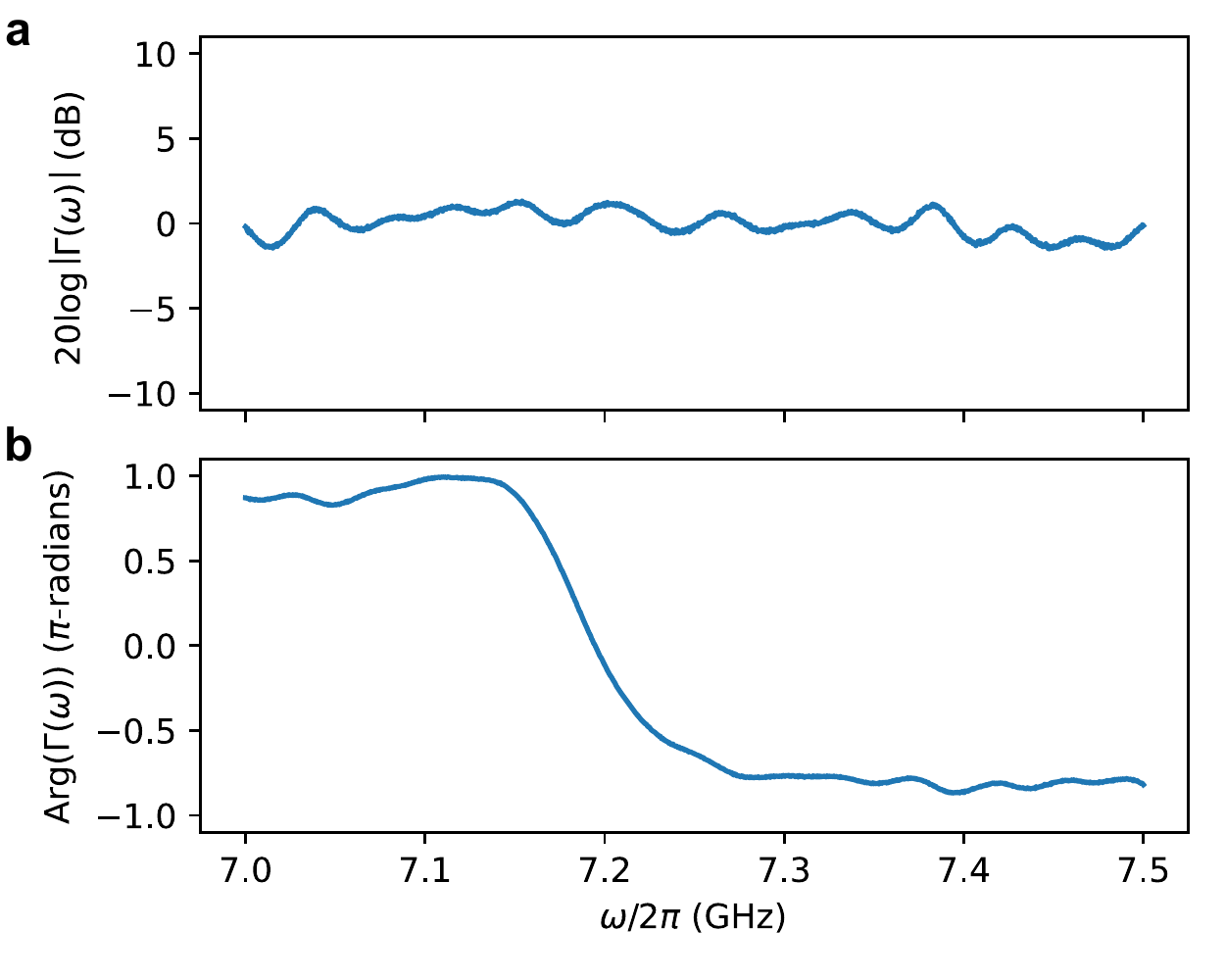}
    \caption{\label{fig:ripple}
        Measured magnitude (a) and phase (b) response of the KIPA investigated in this work. No pump is applied in this measurement and $I_\dc = 0.85$~mA.
    }
\end{figure}

The KIPA operates in the over-coupled regime, where the external coupling rate far exceeds the rate of internal losses ($\kappa \gg \gamma$). As such, the magnitude response in the absence of a pump tone (Fig.~S\ref{fig:ripple}a) is flat, as predicted by input-output theory (\cref{eqn:S11}). We can place a lower bound on the internal quality factor based on the $\sim0.7$~dB ripple observed in our reflection measurement, which indicates $Q_i > 3,350$.

\begin{figure}
    \centering
    \includegraphics[width=0.8\textwidth]{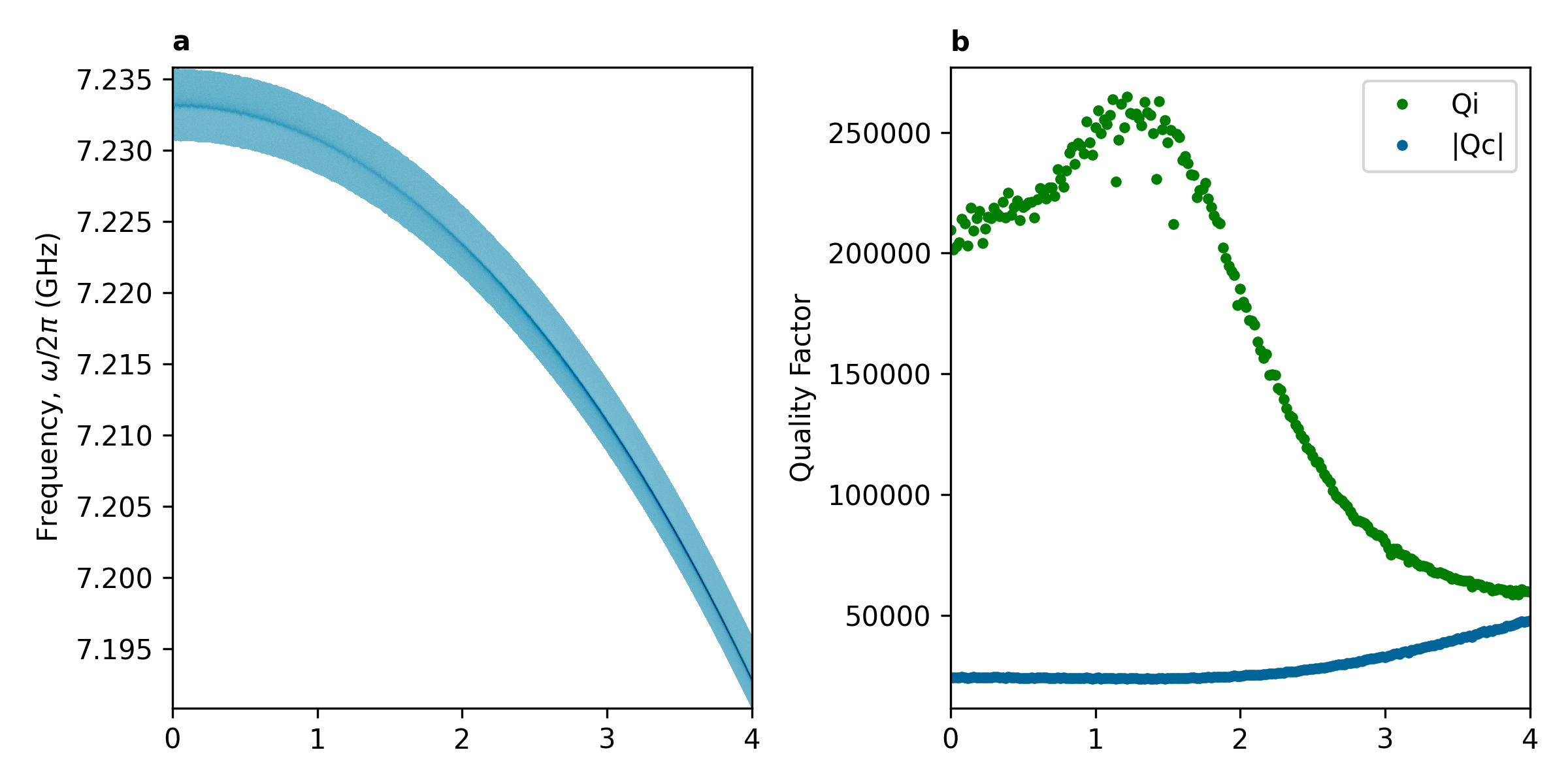}
    \caption{\label{fig:BA13_fits}
        (a) Measured magnitude response of a similar device to the KIPA, as a function of $I_\dc$. (b) Coupling and internal quality factors extracted from fits to the magnitude response in panel a.
    }
\end{figure}

Fig.~S\ref{fig:BA13_fits} depicts the reflection response of a device similar to the KIPA, fabricated on a 50~nm thick NbTiN film and with additional cells in the band stop region to produce a larger external quality factor (i.e smaller $\kappa$). This device operates close to critical coupling where both $\gamma = \omega_0/Q_i$ and $\kappa = \omega_0/Q_c$ may be extracted. Although resonator losses are sensitive to the exact device geometry, this measurement provides an indication of the attainable internal quality factors for Bragg-mirror-coupled microwave resonators. 

We note that the loss in the KIPA will almost certainly depend on its operating conditions. Large intra-cavity fields can induce two photon losses \cite{yurke_2006,eom_2012} and we observe a non-trivial dependence of $Q_i$ on the DC current bias (Fig.~S\ref{fig:BA13_fits}b). Future work will explore the noise properties of the KIPA in further detail, including the search for optimal working points in the device parameter space that maximise noise squeezing.

\bibliography{kipa_SM}
\bibliographystyle{unsrtnat}